\documentclass[journal]{IEEEtran} 
\usepackage{cite}
\usepackage{algorithm,algorithmicx}
\usepackage{algpseudocode}
\usepackage[small,bf]{caption}
\usepackage[cmex10]{amsmath}
\usepackage{amssymb,latexsym,epsfig,subfigure,epic,amscd,mathrsfs,euscript,bm}
\usepackage{color}
\usepackage{booktabs}
\usepackage{array}
\usepackage{amsfonts}
\usepackage{amsopn}
\usepackage{graphicx}
\usepackage{url} 
\usepackage{empheq}
\usepackage{atbegshi}
\usepackage[inline]{enumitem}
\newtheorem{myprop}{\bf{Proposition}}
\newtheorem{theorem}{\bf{Theorem}}
\newtheorem{remark}{\bf{Remark}}

\newcommand{\argmin}{\operatornamewithlimits{arg\,min}}
\newcommand{\argmax}{\operatornamewithlimits{arg\,max}}
\newcommand{\bs}[1]{ \ensuremath{ \boldsymbol{#1} }}
\newcommand{\subscript}[2]{$#1 _ #2$}

\DeclareMathOperator{\mymax}{max}
\DeclareMathOperator{\mymin}{min}
\DeclareMathOperator{\mysup}{sup}
\DeclareMathOperator{\myinf}{inf}

\DeclareMathAlphabet\mathbfcal{OMS}{cmsy}{b}{n}

\begin{document}

\title{Distributed Estimation in Large Scale Wireless Sensor Networks via A Two-Step Cluster-based Approach}


\author{Shan~Zhang, Pranay~Sharma, Baocheng~Geng
	 and Pramod~K.~Varshney,~\IEEEmembership{Life Fellow,~IEEE}	
\thanks{S. Zhang is with the Aptiv PLC, Agoura Hills, CA 91301, USA. Email: \{szhang60\}@syr.edu.}
\thanks{P. Sharma is with the Department
of Electrical and Computer Engineering, Carnegie Mellon University, Pittsburgh, PA 15213, USA. Email: \{pranaysh\}@andrew.cmu.edu.}
\thanks{B. Geng is with the Department
of CS, University of Alabama at Birmingham, Birmingham, AL, 35294, USA. Email: \{bgeng\}@uab.edu.}
\thanks{P. K. Varshney is with the Department of EECS, Syracuse University, Syracuse, NY 13244, USA. Email: \{varshney\}@syr.edu.}
\thanks{This work was supported by National Science Foundation under Grant Engineering 1609916.}
}

\maketitle

\begin{abstract}
We consider the problem of collaborative distributed estimation in a large scale sensor network with statistically dependent sensor observations. In the collaborative setup, the aim is to maximize the overall estimation performance by modeling the underlying statistical dependence and efficiently utilizing the deployed sensors. 
To achieve greater sensor transmission and estimation efficiencies, we propose a two-step cluster-based collaborative distributed estimation scheme. In the first step, sensors form dependence driven clusters such that sensors in the same cluster are dependent while sensors from different clusters are independent, and perform copula-based maximum a posteriori probability (MAP) estimation via intra-cluster collaboration. In the second step, the estimates generated in the first step are shared via inter-cluster collaboration to reach an average consensus. A merge based $K$-medoid dependence driven clustering algorithm is proposed. We further propose a cluster-based sensor selection scheme using mutual information prior to estimation. The aim is to select sensors with maximum relevance and minimum redundancy regarding the parameter of interest under certain pre-specified energy constraint. 
Also, the proposed cluster-based sensor selection scheme is shown to be equivalent to the global sensor selection scheme with high probability, which at the same time is computationally more efficient. Numerical experiments are conducted to demonstrate the effectiveness of our approach.
\end{abstract}

\begin{IEEEkeywords}
Distributed estimation, copula theory, statistical dependence, collaborative estimation, dependence driven clustering, wireless sensor networks, mutual information.
\end{IEEEkeywords}

\section{Introduction}
\label{sec:i}
Wireless sensor networks (WSNs) have attracted significant attention over the past decades due to their high flexibility, robustness, and enhanced coverage. The recent development of low cost, low power and multifunctional sensors has enabled the deployment of large scale WSNs for many applications such as Internet of Things (IoT) and smart cities \cite{gubbi2013internet, jin2014information}.
%
In a distributed estimation problem, sensors are spatially deployed in a large scale network. They sense and collect observations regarding a parameter associated with an object. Sensors are allowed to exchange their information through in-network communication/collaboration, and then transmit their observations or local estimates to a fusion center (FC) which produces a global estimate. Since sensors are typically battery operated, and located far away from the FC, it is expensive to transmit their raw observations or local estimates to the FC. Therefore, in this paper, we study a fully distributed estimation problem, where there is no FC, and sensors collaborate under certain pre-specified protocols and estimate the parameter of interest.

In large scale sensor networks, sensor observations can be dependent or independent. Distributed estimation problems with independent sensor observations have been studied extensively (see e.g. \cite{luo2005isotropic, fang2008distributed,fang2009power2, kar2013linear, liu2014sparsity, schizas2009distributed, cattivelli2010diffusion, kar2010distributed, chiuso2011gossip}). However, handling of dependent sensor observations is a critical issue in distributed estimation.  The underlying dependence can be both good and bad \cite{yoon1999effect, he2013coalitional, he2016coalitional}. On the one hand, dependent sensors provide different viewpoints and aspects regarding the target parameter to be estimated. However, on the other hand, they may collect redundant observations. Therefore, spatial dependence needs to be exploited properly to enhance the overall estimation efficiency. In \cite{he2013coalitional, he2016coalitional}, the concepts of diversity gain and redundancy loss were introduced to characterize the influence of spatial dependence among sensor observations on estimation performance. 

In the presence of sensor collaboration, the problem of distributed estimation has attracted significant attention (see e.g. \cite{fang2009power2, kar2013linear, liu2014sparsity, schizas2009distributed, cattivelli2010diffusion, kar2010distributed, chiuso2011gossip} and references therein).
In \cite{fang2009power2, kar2013linear, liu2014sparsity}, distributed estimation problems with a  FC were considered, where collaboration was restricted to be a linear operation. Moreover, the collaboration in \cite{fang2009power2} was conducted within pre-specified sensor clusters and the clusters were assumed to be independent. Collaborative distributed estimation problems without the FC were studied in \cite{schizas2009distributed, cattivelli2010diffusion, kar2010distributed, chiuso2011gossip}, where different distributed collaboration strategies were proposed, such as diffusion-based, consensus-based and gossip-based algorithms. However, most of the works assume independent observations across sensors, and much less has been done for the dependent observations case. Ignoring dependence may not fully exploit the information available from the observations. 
Therefore, in this paper, we take the underlying dependence among sensor observations into account and seek an efficient estimation scheme.

The problem of distributed estimation with dependent observations has attracted some recent attention.  
In the absence of collaboration, linearly dependent observation noise was taken into account in \cite{fang2009power}. In \cite{sundaresan2011location}, a copula based approach was proposed at the fusion center to characterize the non-linear dependence among sensor observations and improve the overall estimation performance. The copula based approach \cite{joe2014dependence} is a flexible parametric dependence modeling methodology, where the  joint distribution of multiple sensor observations can be modeled using marginal distributions and a multivariate (dependence) distribution, which is referred to as a multivariate copula. In the presence of collaboration, in \cite{he2013coalitional, he2016coalitional}, a collaborative copula based distributed estimation scheme was proposed for large scale sensor networks, where spatial dependence was exploited to maximize the performance of non-overlapping coalitions under certain energy constraints, and non-linear sensor collaboration was conducted within coalitions.  

In this paper, we propose a novel two-step collaborative distributed estimation framework, taking into account the inherent tradeoff between estimation performance and transmission efficiency. 
In such a framework, each sensor is able to sense the parameter of interest and perform estimation. More specifically, sensors estimate the target parameter in the first step, and share their estimates based on certain pre-defined network communication topology in the second step. Since the size of the network is potentially large, it is necessary to limit the amount of information flowing through the network and avoid unneeded power consumption. It has been shown that cluster-based topologies are less energy consuming than regular non-cluster based network topologies \cite{lloret2008improving}. Therefore, in our framework, to achieve greater transmission efficiency, we allow cluster-based collaboration, and incorporate two types of collaboration: intra-cluster communication in the first step and inter-cluster communication in the second step (see Fig. \ref{fig:system_group}). 

Motivated by the fact that sensor observations in the network can be dependent as well as independent, 
we assume that sensors in the network can be clustered into independent non-overlapping clusters. The details of the two-step cluster-based collaborative distributed estimation framework are given as follows. In the first step, sensors form dependence driven non-overlapping clusters. We propose a merge based $K$-medoid dependence driven clustering algorithm. Once the clusters are formed, via intra-cluster collaboration, each sensor estimates the target parameter of interest using copula based MAP which takes into account the non-linear dependence among the sensors in the same cluster.  In the second step, the estimates obtained in the first step are shared among clusters until an average consensus is reached.


In large scale sensor networks, efficient management and utilization of the deployed sensors is an important consideration. Since sensors that are in the same cluster are dependent, some sensors may provide redundant information. For the purposes of efficient utilization of the deployed sensors and increasing the lifetime of WSNs, one typically designs policies to optimally select a subset of sensors that are most informative and least redundant. In this way, we can reduce the intra-cluster transmission cost by exploiting the spatial dependence across selected sensors within each cluster and reduce the computational complexity resulting from the underlying dependence. In this work, we propose a sensor selection incorporated two-step cluster-based collaborative estimation scheme. More specifically, 
each cluster uses a mutual information based scheme to select sensors with maximum information and minimum redundancy under certain energy cost constraints before estimating the target parameter of interest. After that, the selected sensors collaborate within clusters to perform estimation.

The differences between this paper and the work in \cite{he2013coalitional, he2016coalitional, schizas2009distributed, cattivelli2010diffusion, kar2010distributed, chiuso2011gossip} are stated as follows. We first present the differences with respect to \cite{he2013coalitional, he2016coalitional}. 1), the metric employed for clustering of the sensors is different. In \cite{he2013coalitional, he2016coalitional}, the coalitions/clusters were determined by maximizing the total average Fisher information which improved estimation performance, while in this paper, the clusters are formed by maximizing the total dependence/similarity in such a way that dependent sensors are included in the same cluster. 2), the collaboration scheme is different. In \cite{he2013coalitional, he2016coalitional}, each sensor performed estimation via intra-coalition collaboration. In contrast, in this paper, we allow two types of collaboration, namely, intra-cluster and inter-cluster collaboration. 
3), the estimation scheme is different. In \cite{he2013coalitional, he2016coalitional}, sensors used maximum likelihood estimation (MLE) to estimate the parameter of interest, where the underlying non-linear dependence was not addressed. In this paper, we propose a copula based MAP approach for estimation. Compared to the work in \cite{schizas2009distributed, cattivelli2010diffusion, kar2010distributed, chiuso2011gossip} where individual sensor estimates were obtained independently and combined based on linear collaboration, we take dependent sensor observations into account, and the initial estimates of the target parameter of interest are based on intra-cluster collaboration which is more robust and reliable compared to the individual sensor estimation. Also, intra-cluster collaboration considered here is non-linear, where the underlying dependence among sensor observations is utilized.
We summarize our contributions as follows. 
\begin{itemize}
\item We take non-linear dependence among sensor observations into account for the distributed estimation problem in large scale sensor networks and propose a two-step cluster-based collaborative distributed estimation scheme. 
\item We propose a merge based $K$-medoid dependence driven algorithm for the clustering of sensors. 
\item We propose a copula based MAP approach in each cluster to estimate the target parameter of interest.
\item We propose an inter-cluster average consensus scheme, and we show that the standard deviation of the estimate obtained by the average consensus method is upper bounded by the average standard deviation of the cluster estimates.
\item We propose a sensor selection based two-step cluster-based collaborative distributed estimation scheme to select the most informative and least redundant sensors in each cluster. 
\item We show that cluster-based sensor selection methodology is equivalent to the global sensor selection method with high probability.
\item We show the superiority of our proposed two-step cluster-based collaborative estimation methodologies via a number of illustrative examples.  
\end{itemize}

The rest of the paper is organized as follows. In Section \ref{sec:CT}, we provide a brief introduction to copula theory. In Section \ref{sec:PF}, we introduce the two-step cluster-based collaborative distributed estimation system, and state the distributed estimation problem. In Section \ref{sec:twostep}, we present the details of our proposed estimation scheme including the merge based $K$-medoid dependence driven clustering algorithm, the copula-based MAP estimation scheme and the inter-cluster consensus scheme. In Section \ref{sec:sensorselection}, we propose a cluster-based sensor selection strategy for the two-step collaborative distributed estimation system.  In Section \ref{sec:nr}, we demonstrate the effectiveness of the proposed estimation scheme through numerical examples. Finally, in Section \ref{sec:conclusion}, we summarize our work and discuss future research directions.

\section{Copula Theory}
\label{sec:CT}
A copula is a multivariate distribution with uniform marginal distributions, and it characterizes the dependence among multiple continuous variables. 
The unique correspondence between a multivariate copula and any multivariate distribution is stated in Sklar's Theorem \cite{nelsen2013introduction} which is a fundamental theorem that forms the basis of copula theory. 

\begin{theorem}[Sklar's Theorem]
The joint distribution function $F$ of random variables $x_1,\ldots,x_d$ can be cast as
\begin{equation}
\label{CopEq1}
F(x_1,x_2,\ldots,x_d) = C(F_1(x_1),F_2(x_2),\ldots,F_d(x_d) | \bs \phi),
\end{equation}
where $F_1,\ldots,F_d$ are marginal distribution functions for $x_1,\ldots,x_d$.  If $F_m, m = 1, \ldots, d$ are continuous, $C$ is a unique $d$-dimensional copula with \textit{dependence parameter} $\bs{\phi}$. Conversely, given a copula $C$ and univariate Cumulative Distribution Functions (CDFs) $F_1,\ldots,F_d$, $F$ in Equation \eqref{CopEq1} is a valid multivariate CDF with marginals $F_1,\ldots,F_d$. Note that $\bs{\phi}$ is used to characterize the amount of dependence among the $d$ random variables. In general, $\bs{\phi}$ may be a scalar, a vector or a matrix
\end{theorem}

For continuous distributions $F$ and $F_1,\ldots,F_d$, the joint Probability Density Function (PDF) of random variables $x_1,\ldots,x_d$ is obtained by differentiating both sides of Equation \eqref{CopEq1}:
\begin{equation}
\label{CopEq2}
f(x_1,\ldots,x_d) \!=\! \Big(\!\!\prod_{m=1}^{d}f_m(x_m)\!\Big)c(F_1(x_1),\ldots,F_d(x_d) | \bs \phi),
\end{equation}
where $f_1, \ldots, f_d$ are the marginal densities and $c$ is referred to as the density of the multivariate copula $C$ that is given by 
\begin{equation}
c(\mathbf{u} | \bs \phi) = \frac{\partial^d(C(u_1,\ldots,u_d | \bs \phi))}{\partial u_1,\ldots,\partial u_d},
\end{equation}
where $u_m=F_m(x_m)$ and $\mathbf{u} = [u_1,\dots,u_d]$. 
Note that $C(\cdot)$ is a valid CDF and $c(\cdot)$ is a valid PDF for uniformly distributed random variables $u_m$, $m = 1, 2, \ldots, d$. Since the random variable $u_m$ represents the CDF of $x_m$, the CDF of $u_m$ naturally follows a uniform distribution over $[ 0,1]$.


Various families of multivariate copula functions are presented in \cite{nelsen2013introduction}, such as elliptical and Archimedean copulas. Since different copula functions model different types of dependence, selection of copula functions to fit the given data is a key problem. 
Moreover, the dependence parameter $\bs{\phi}$ is typically unknown a \textit{priori} and needs to be estimated, e.g., using MLE or Kendall's $\tau$ \cite{he2015heterogeneous}.

\section{Problem Formulation}
\label{sec:PF}
Consider a phenomenon being observed by $L$ sensors. Each sensor's observation is $z_l = \theta + w_l, \forall l = 1, \ldots, L$, where $\theta$ is the random parameter to be estimated corresponding to the phenomenon of interest and $w_l$ is the observation noise which is spatially and temporally independent of $\theta$. We assume that the prior distribution of $\theta$ is given as $f(\theta)$. Also, we assume that the observation noise can be dependent across some sensors. Moreover, we further assume that the sensor observations are continuous random variables that are conditionally independent and identically distributed (i.i.d.) over time. Let $f_l(\cdot | \theta)$ be the PDF of the observations at the $l$th sensor conditioned on $\theta$. Note that the marginal conditional sensor PDFs can be distinct from each other. Throughout the paper, we assume that given $\theta$, the marginal distribution $f_l(\cdot | \theta), l = 1, \ldots, L$ is known. 

\begin{figure}
\centering
\includegraphics[height=2.4in,width=!]{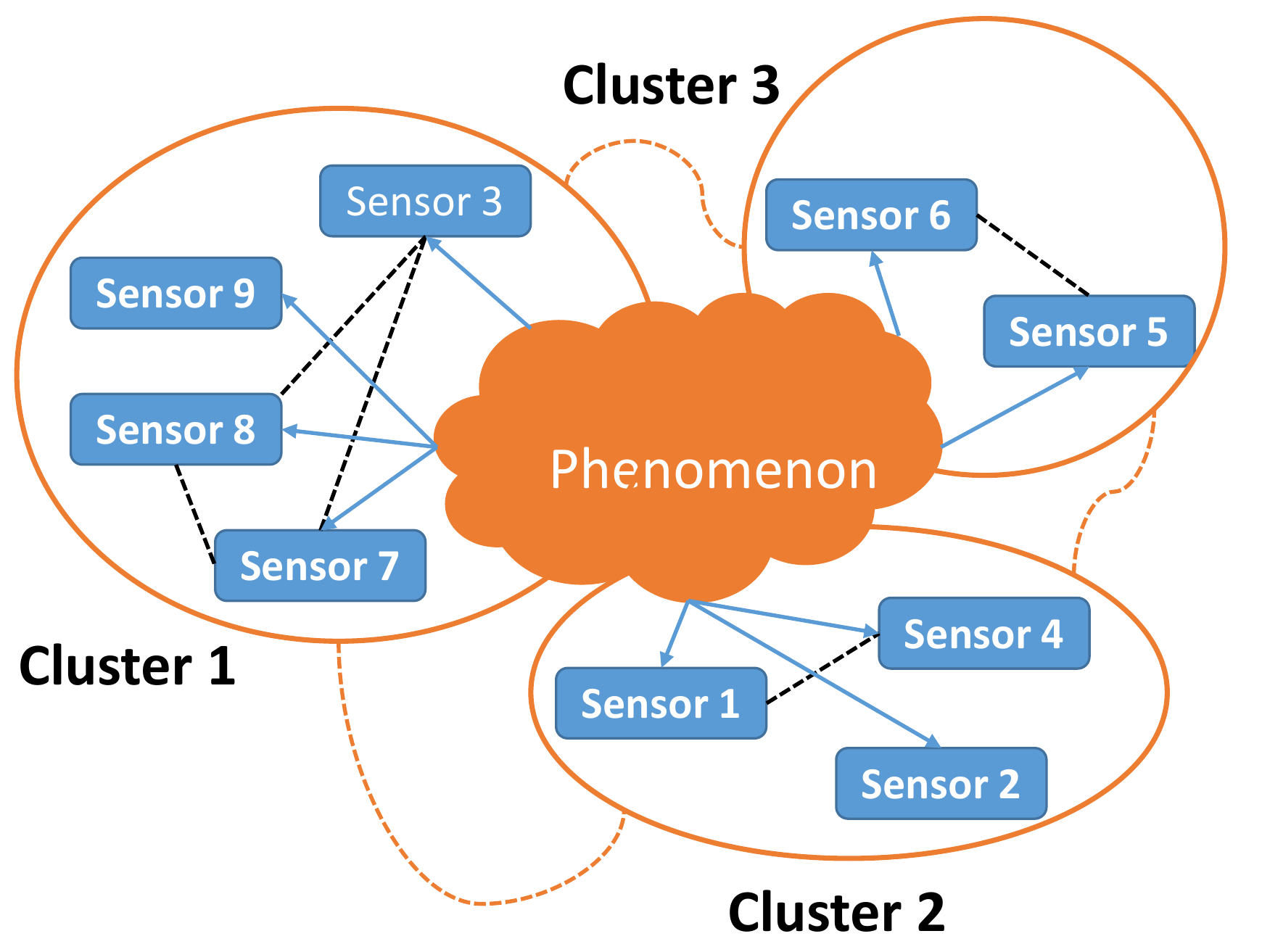}
\caption{Two-step cluster-based collaborative distributed estimation system, where the orange dash lines represent the inter-cluster communication links and the black dash lines denote the intra-cluster communication links.}
\label{fig:system_group}
\end{figure}

In a non-collaborative setting, each sensor senses the phenomenon of interest and estimates the random parameter $\theta$ solely based on its own observations. In this work, we consider a two-step cluster-based collaborative scheme shown in Fig.\,\ref{fig:system_group}, where in the first step, sensors form dependence driven clusters and 
extract information relevant for estimation by collaborating with other sensors in the same cluster. In the second step, local information obtained by each cluster in the first step is shared among clusters to yield a global estimate. 
The participating sensors are required to adhere to the following rules:
\begin{enumerate}
\item Sensors first form clusters, where each sensor is allowed to join only one cluster. The sensors that are most ``similar", i.e., most statistically dependent, tend to stay in the same cluster.
\item Once the clusters are formed, a sensor can request observations from all the other sensors that are in the same cluster to perform estimation; 
it is also required to transmit its observations to the other collaborating sensors in the cluster based on their request.
\item A cluster can request the estimate of the parameter or observations from all the other clusters; it is also required to transmit its estimate of the parameter or observations to the other collaborating clusters based on their request.
\end{enumerate}

We denote the set of all the sensors in the network as $\mathcal S$, where the corresponding sensor observation set is $\mathbf z_{\mathcal S} = [\mathbf z_1, \ldots, \mathbf z_{|\mathcal S|}] \in \mathbb R^{N \times |\mathcal S|}, |\mathcal S| = L$, where $|\cdot|$ denotes the cardinality of a set and $N$ is the number of observations for each sensor. Suppose there are $K$ independent non-overlapping sensor clusters and denote the $k$th cluster by $ \mathcal G_k, k \in [K]$, where for ease of notation, $[K]$ denotes $\{1, 2, \ldots, K\}$. Thus, $\mathcal S =  \mathcal G_1 \cup \cdots \cup \mathcal G_{K}$.

In the estimation problem, Fisher Information (FI) is often used to characterize the amount of information that data carry about the parameter. It is given as
\begin{equation}
FI(\theta) = - \mathbb E_{\mathbf x} \left[ \frac{\partial^2 \text{log} f_{\mathbf x}(\mathbf x; \theta)}{\partial \theta^2} \right],
\end{equation}
where $f_{\mathbf x}$ represents the joint PDF of the data sequence vector $\mathbf x$.  For the entire sensor set $\mathcal S$, the FI it can achieve is given as
\begin{equation}
FI(\mathcal S) = - \mathbb E_{\mathbf z_{\mathcal S}} \left [ \frac{\partial^2 \text{log} f_{\mathbf z_{\mathcal S}}(\mathbf z_{\mathcal S}; \theta)}{\partial \theta^2}  \right],
\end{equation}
where $f_{\mathbf z_{\mathcal S}}$ is the joint distribution of $\mathbf z_{\mathcal S}$.

\begin{myprop}
\label{prop:fi}
Since we assume that sensors in the network can be clustered into independent
non-overlapping clusters, $FI(\mathcal S)$ can be decomposed into  cluster-based Fisher Information and prior Fisher Information.
\end{myprop}
\textbf{Proof}: \begin{align*}
&FI(\mathcal S) \\ \nonumber
&= - \mathbb E \left[ \frac{\partial^2 \text{log} \left ( \prod_{l=1}^{L}f_l(\mathbf z_l | \theta) \times c_{\mathcal S}(\mathbf F(\mathbf z | \theta) ; \bs \phi) \times f(\theta) \right )}{\partial \theta^2} \right], \\ \nonumber
&= \sum_{l \in \mathcal S}FI_{l} + FI_p - \mathbb E \left [   \frac{\partial^2 \text{log} \, c_{\mathcal S}(\mathbf F(\mathbf z | \theta) ; \bs \phi)}{\partial \theta^2} \right], \\ \nonumber
&\overset{(a)}{=} \sum_{l \in \mathcal S}FI_{l} + FI_p - \mathbb E \left [   \frac{\partial^2 \text{log} \, \prod_{k=1}^{K} c_{\mathcal \mathcal G_k}(\mathbf F(\mathbf z_{\mathcal{G}_k} | \theta) ; \bs \phi_k)}{\partial \theta^2} \right], \\ \nonumber
&= \sum_{l \in \mathcal S}FI_{l} + \sum_{k=1}^K FI_c(\mathcal G_k) + FI_p, \nonumber \\
&= \sum_{k=1}^K \left ( \sum_{l \in \mathcal G_k}FI_{l} + FI_c(\mathcal G_k) \right) + FI_p, \nonumber
\end{align*}
where $FI_p$ is the Fisher information with respect to the prior distribution on $\theta$,  $(a)$ is obtained by using the assumption that sensor clusters are independent of each other. Also, we define $FI_c(\mathcal G_k)$ as $-\mathbb E \left [   \frac{\partial^2 \text{log} \, c_{\mathcal \mathcal G_k}(\mathbf F(\mathbf z_{\mathcal{G}_k} | \theta) ; \bs \phi_k)}{\partial \theta^2} \right]$.

Therefore, $FI(\mathcal S)$ can be decomposed into cluster-based Fisher Information and prior Fisher Information. \hfill $\blacksquare$

\begin{remark}
Based on Proposition \ref{prop:fi}, we can process each cluster independently and then combine each cluster's information to obtain the global estimate.
\end{remark}

In the first step, an intuitive solution would be that each cluster learns its dependence structure, and shares the estimated conditional joint PDFs with all the other clusters in the second step.  Then,  the estimation problem becomes
\begin{align}
\label{eq:joint2}
\hat{\theta} = \underset{\theta}{\text{arg max}} &\sum_{i=1}^{N} \sum_{k=1}^{K} \left (\text{log}\, f(\mathbf z_{\mathcal G_k, i} | \theta) \right) + \text{log}\, f(\theta), 
\end{align}
where $\mathbf z_{\mathcal G_k, i}$ is the observation set for cluster $\mathcal G_k$ at time instant $i$ and $f(\mathbf z_{\mathcal G_k} | \theta)$ is the conditional joint PDF of the sensor observations (which is not known a \textit{priori}) in cluster $\mathcal G_k, k \in [K]$. $f(\theta)$ is the prior distribution of $\theta$.

\begin{remark}
The estimation methodology given in Equation \eqref{eq:joint2} is referred to as \textit{cluster-based MAP} scheme. Note that the conditional joint PDF $f(\mathbf z_{\mathcal G_k} | \theta)$ in Equation \eqref{eq:joint2} can be estimated using copula based methods that take dependent observations into consideration (see Equation \eqref{CopEq2}). The cluster-based MAP scheme using copula incorporated approach is optimal.
\end{remark}

However, transmitting the estimated conditional joint PDFs and the raw observations among clusters can be expensive. Therefore, we propose to share estimates obtained by each cluster until a consensus is achieved. 

In the following section, we present the details of our two-step cluster-based distributed estimation scheme, including the clustering of the sensors, the intra-cluster collaborative estimation approach using copula based methods and the inter-cluster collaboration strategy. 

\section{Two-Step Dependence Driven Collaborative Distributed Estimation}
\label{sec:twostep}
In this section, we present our two-step cluster-based collaborative distributed estimation scheme. In the first step, sensors form clusters based on their similarity/dissimilarity with the other sensors. We propose a merge based $K$-medoid dependence driven clustering algorithm. After the clusters are formed, each sensor then estimates $\theta$ using copula based MAP via intra-cluster collaboration. In the second step, the estimated $\theta$s are shared among clusters to yield a consensus.  Here, we assume that the sensors and the sensor clusters communicate via error-free, orthogonal channels. Before we proceed, we first make some assumptions and define the dissimilarity measures.

\subsection{Assumptions and Dissimilarity Measure Definitions}
We define the inter-cluster dissimilarity between $\mathcal G_k$ and $\mathcal G_{k^{\prime}}$ as well as the intra-cluster dissimilarity of $\mathcal G_k$, respectively as
\begin{align*}
d(\mathcal G_k, \mathcal G_{k^{\prime}}) &=  \underset{s_i \in \mathcal G_k, s_j \in \mathcal G_{k^{\prime}}}{\myinf}  \mathit d(s_i, s_j), \\
d(\mathcal G_k) &= \underset{s_i, s_j \in \mathcal G_k}{\mysup} \mathit d(s_i, s_j), 
\end{align*}
where $d(\cdot, \cdot)$ is a dissimilarity metric between two variables/data sequences, e.g., the rank based dissimilarity measure defined later in Equation \eqref{eq:d1}. Here, $d(\mathcal G_k, \mathcal G_{k^{\prime}})$ represents the dissimilarity between cluster $\mathcal G_k$ and $\mathcal G_{k^{\prime}}$. 
We further define
\begin{align*}
d_H &= \underset{k, k^{\prime} = 1, \ldots, K, k \neq k^{\prime}}{\mymin} \,  d(\mathcal G_k, \mathcal G_{k^{\prime}}), \\
d_L &= \underset{k = 1, \ldots, K}{\mymax} \, d(\mathcal G_k).
\end{align*}

We make the following assumptions:
\begin{enumerate}[label=\subscript{A}{{\arabic*}}]
\item $d_L < d_H$,
\item $P\left (d(\mathbf z_i^{k}, \mathbf z_j^{k^{\prime}}) \leq d_0 \right) < \epsilon_1, d_0 \in (d_L, d_H)$,
\item $P\left (d(\mathbf z_i^{k}, \mathbf z_j^{k}) > d_0 \right) < \epsilon_2, d_0 \in (d_L, d_H)$,
\item $P\left (d(\mathbf z_i^{k}, \mathbf z_j^{k}) \geq d(\mathbf z_i^{k}, \mathbf z_{j^{\prime}}^{k^{\prime}}) \right) < \epsilon_3$,
\end{enumerate}
where $\mathbf z_i^{k}, \mathbf z_j^{k} \in \mathbf z_{\mathcal G_k}$ and $\mathbf z_j^{k^{\prime}}, \mathbf z_{j^{\prime}}^{k^{\prime}} \in \mathbf z_{\mathcal G_{k^{\prime}}}$. $\epsilon_i > 0, i = 1, 2, 3$ are small constants and $N$ is the number of observations.

Assumption $A_2$ implies that the probability that the dissimilarity between sensor observations obtained from two different clusters is smaller than $d_H$ is small. Also, assumption $A_3$ guarantees that the probability that the dissimilarity between sensor observations obtained from the same cluster is greater than $d_L$ is small. Assumption $A_4$ states that given two sensor observation sequences generated from the same cluster and a third observation sequence generated from another cluster, the probability that the first sequence is closer to the third sequence is small. 
Due to the use of measured noisy data, the assumptions $A_1$ to $A_4$ imply that sensors that are from the same cluster are dependent, while the ones that are from different clusters are nearly independent.

The dissimilarity between two sensors can be characterized using different dependence measures, such as the Pearson's correlation coefficient, a rank based correlation measure (Spearman's $\rho$ and Kendall's $\tau$) and the copula based measure. Note that the Pearson's correlation coefficient that characterizes linear relationship is inadequate to capture nonlinear dependence among the involved sensors. Also, the copula based measure is not a symmetric dependence measure. In the following, we propose a dissimilarity metric based on rank based correlation. 

Let $\kappa \in [-1, 1]$ be a rank based measure (Spearman's $\rho$ or Kendall's $\tau$). We introduce a dissimilarity function $d(\mathbf x, \mathbf y)$ between the random variables $X$ and $Y$, where $\mathbf x = [x_1, \ldots, x_N]$ and $\mathbf y = [y_1, \ldots, y_N]$ are the i.i.d. data sequences corresponding to the variables $X$ and $Y$, respectively, given as
\begin{equation}
\label{eq:d1}
\mathit d(\mathbf x, \mathbf y) = \sqrt{1 - \kappa(\mathbf x, \mathbf y)^2},
\end{equation}
where $N$ is the number of samples for variables $X$ and $Y$ and $\kappa(\mathbf x, \mathbf y)$ is Spearman's $\rho$ or Kendall's $\tau$ between sequences $\mathbf x$ and $\mathbf y$. Note that if $\kappa(\mathbf x, \mathbf y) = 1$ or $\kappa(\mathbf x, \mathbf y) = -1$, we have $\mathit d (\mathbf x, \mathbf y) = 0$.

\subsection{Dependence Driven Clustering Process}
We propose a dependence driven clustering scheme.  
Let $d(\mathbf z_i, \mathbf z_j)$ denote the dissimilarity between the $i$th and $j$th sensors, where $i, j \in [L]$. Therefore, $d(\mathbf z_i, \mathbf z_j)$ is small when sensor $i$ and sensor $j$ are strongly dependent and is large when sensor $i$ and sensor $j$ are weakly dependent.

 The goal of the clustering process is to cluster the sensors in the network based on the underlying dependence among sensors. 
The number of clusters $K$ is unknown. Therefore, we need to estimate it. Here, we apply a merge based $K$-medoid clustering scheme \cite{xiong2004similarity, apt2009generic, he2013coalitional, he2016coalitional} to perform the clustering and find $\hat{K}$. The merging criterion is that if the dissimilarity/distance of any two clusters is greater than $d_{th} \in (d_L, d_H)$, these two clusters should be separated; otherwise, they merge together. 

In the following, we present the initialization of the cluster centers and clusters. Before, we initialize the clusters, the centers need to be initialized first. We denote the cluster centers as $\bs {\mu}_1, \ldots, \bs {\mu}_{\hat K}$ and the cluster center set as $\bs \mu$.
 We first arbitrarily choose $\mathbf z_i, i \in [L]$ as  $\bs {\mu}_1$ and $\bs \mu = \{ \bs {\mu}_1 \}$. Then, for
$\underset{\mathbf z_i \in \mathbf z_{\mathcal S} \setminus \bs \mu}{\mymax} \, \left (\underset{\bs {\mu}_k \in \bs \mu}{\mymin} \, d \left ( \mathbf z_i, \bs {\mu}_k  \right ) \right) > d_{th}$, we do
\begin{align}
&\tilde {\bs {\mu}} = \underset{\mathbf z_i \in \mathbf z_{\mathcal S} \setminus \bs \mu}{\argmax} \, \left (\underset{\bs {\mu}_k \in \bs \mu}{\mymin} \, d \left ( \mathbf z_i, \bs {\mu}_k  \right ) \right), \\ \nonumber
&\bs \mu = \bs \mu \cup \tilde {\bs {\mu}}.
\end{align}

After we obtain the cluster centers, the clusters, which are originally defined as empty sets,  are initialized as: for  $i = 1, 2, \ldots, L$
\begin{align}
&\bs \mu_j = \underset{\mathbf \mu_j \in \bs \mu}{\argmin} \, d\left ( \mathbf z_i, \bs \mu_j \right ), \\ \nonumber
&\mathcal G_j \gets \mathcal G_j \cup \{ \mathbf z_i \}. \nonumber
\end{align}

The proposed dependence driven clustering scheme is shown in Algorithm \ref{algo:cluster}. 
\begin{algorithm}
	\textbf{Input:} Sensor observations $\{ \mathbf z_1, \ldots, \mathbf z_L\}$ and threshold $d_{th}$.
	
	\textbf{Output:} Clusters $\{ \mathcal G_k \}_{k=1}^{\hat{K}}$.
	\begin{enumerate}
	         \item Initialize clusters $\{ \mathcal G_k \}_{k=1}^{\hat{K}}$
	         \item \textbf{while} not converge \textbf{do}
	         \item \textbf{Center update:}
	         \item \quad \textbf{for} $k = 1$ to $\hat{K}$ \textbf{do}
	         \begin{equation*}
	         \bs {\mu}_k \gets \underset{\mathbf z_l \in \mathcal G_k}{\text{arg min}} \, \sum_{\mathbf z_{l^{\prime}} \in \mathcal G_k}\, d\left (\mathbf z_l, \mathbf {z}_{l^{\prime}} \right) 
	         \end{equation*}
	       \item  \quad \textbf{end for}
	       \item \textbf{Merge step:}
	       \item  \quad\textbf{for} $k_1, k_2 \in [1, 2, \ldots, \hat{K} ]$ and $k_1 \neq k_2$ \textbf{do}
	       \item  \quad  \quad \textbf{if} $d(\bs \mu_{k_1}, \bs \mu_{k_2}) \leq d_{th}$  \textbf{then} 
	       \item  \quad  \quad  \quad  \quad \textbf{if} $\sum_{\mathbf z_l \in \mathcal G_{k_1}} d(\bs \mu_{k_2}, \mathbf z_l) < \sum_{\mathbf z_l \in \mathcal G_{k_2}} d(\bs \mu_{k_1}, \mathbf z_l)$
	       \item  \quad  \quad  \quad  \quad \textbf{then}
	       \begin{equation*}
	      \quad  \quad  \quad  \quad \mathcal G_{k_2} \gets \mathcal G_{k_1} \cup \{ \mathcal G_{k_2}  \} \,\, \text{and delete} \,\, \bs \mu_{k_1}\,\, \text{and}\,\, \mathcal G_{k_1}
	       \end{equation*}
	       \item  \quad  \quad  \quad  \quad \textbf{else}
	       	       \begin{equation*}
	       \quad  \quad  \quad  \quad \mathcal G_{k_1} \gets \mathcal G_{k_1} \cup \{ \mathcal G_{k_2}  \} \,\, \text{and delete} \,\, \bs \mu_{k_2}\,\, \text{and}\,\, \mathcal G_{k_2}
	       \end{equation*}
	       \item  \quad  \quad  \quad  \quad \textbf{end if}
	       \item $\hat{K} \gets \hat{K} - 1$
	       \item  \quad  \quad \textbf{end if}
	       \item  \quad \textbf{end for}
	       \item \textbf{Cluster update:}
	       \item \quad \textbf{for} $l = 1$ to $L$ \textbf{do}
	       \item \quad \quad \textbf{if} $\mathbf z_l \in \mathcal G_{k^{\prime}}$ and $d\left (\mathbf z_l, \bs {\mu}_k \right) <  d\left (\mathbf z_l, \bs {\mu}_{k^{\prime}} \right)$ then
	       \begin{equation*}
	       \mathcal G_k \gets \mathcal G_k \cup \{ \mathbf z_l \} \,\, \text{and} \,\, \mathcal G_{k^{\prime}} \gets \mathcal G_{k^{\prime}} \setminus \{ \mathbf z_l \}
	       \end{equation*}
	       \item \quad \quad \textbf{end if}
	       \item \quad \textbf{end for}
	       \item \textbf{end while}
	       \item \textbf{Return} $\{ \mathcal G_k \}_{k=1}^{\hat K}$
		\end{enumerate}
	\caption{Dependence Driven Clustering.}
	\label{algo:cluster}
\end{algorithm}

\subsection{Copula Based MAP}
After the clusters are formed, each sensor performs estimation by collaborating with the sensors in the same cluster. We assume a fully connected network for intra-cluster collaboration. In each cluster, each sensor estimates $\theta$ using MAP based on its own observations and observations from all the other collaborating sensors in the same cluster. Note that for a fully connected network, all the sensors in the cluster have the same set of observations available to them. 
We denote the corresponding sensor observations for cluster $\mathcal G_k$ as $\mathbf z_{\mathcal G_k} = \{\mathbf z_{1}^k, \mathbf z_{2}^k, \ldots, \mathbf z_{|\mathcal G_k|}^k \}$. 
Therefore, the estimate $\hat{\theta}_k$ at each sensor for the $k$th cluster is given by
\begin{equation}
\hat{\theta}_k = \underset{\theta}{\text{arg max}} \sum_{i=1}^{N} \text{log} \left (f(z_{1i}^k, z_{2i}^k, \ldots, z^k_{|\mathcal G_k|i}| \theta) \times f(\theta) \right),
\end{equation} 
where $N$ is the number of observations and $f(z_{1i}^k, z_{2i}^k, \ldots, z_{|\mathcal G_k|i}^k; \theta)$ is the joint PDF of all the sensors in cluster $\mathcal G_k, k \in [K]$ at time instant $i, i \in [N]$.

We use the copula based approach to characterize the underlying dependence in each cluster and according to Equation \eqref{CopEq2}, $\hat{\theta}_k$ can be obtained by
\begin{align}
\label{eq:joint}
\hat{\theta}_k = \underset{\theta}{\text{arg max}} &\sum_{i=1}^{N} \sum_{l=1}^{|\mathcal G_k|} \text{log}\, f_l(z_{li}^k | \theta) + \sum_{i=1}^{N} \text{log}\, c_k(\mathbf F(\mathbf z_i^k| \theta) ; \bs \phi_k) \\ \nonumber
&+ \text{log}\, f(\theta), \nonumber
\end{align}
where $\mathbf F(\mathbf z_i^k | \theta) =  [F(z_{1i}^k | \theta), F(z_{2i}^k | \theta), \ldots, F(z_{|\mathcal G_k|i}^k | \theta)]$ is the set of marginal CDFs, and $c_k(\cdot ; \bs \phi_k)$ is the multivariate copula density function and $\bs{\phi}_k$ is the corresponding parameter set for cluster $k, k \in [K]$.

Typically,  the multivariate dependence $c_k(\cdot; \bs \phi_k)$ in Equation \eqref{eq:joint} is unknown a \textit{priori} and needs to be estimated. Since the learning of the copula models is similar for all the clusters, in the following, we omit the cluster index $k$ for simplification of notation. 

To estimate the multivariate copula $c(\cdot; \bs \phi)$, we first define a library of copula models, $\mathcal{C} =  \{c_{j}: j =1,\ldots,M\}$ \cite{nelsen2013introduction}. The optimal copula model is then determined by the Akaike Information Criterion (AIC) \cite{akaike1973information} in Equation \eqref{eq:aiccc}, namely, the best copula is the copula model with minimum AIC value. Before evaluating the AIC values for each copula model, we need to estimate the marginal CDFs and the associated copula parameter(s) $\bs \phi_{j}, j = 1, \ldots, M$.  The marginal CDFs can be estimated using EPIT \cite{he2012fusing}:
\begin{equation}
\hat{F}_l(x) = \frac{1}{N} \sum_{n=1}^{N} \bs{\mathnormal{I}}(z_{ln} < x),
\label{empCDFff}
\end{equation}
where $\bs{\mathnormal{I}}$ is the indicator function and $N$ is the number of observations for estimation. The copula parameter(s) $\bs{\phi}_{j}$ can then be estimated using MLE, which is given by
\begin{equation}
\label{eq:phiii}
\widehat{\bs{\phi}}_{j} = \arg\max_{ \bs{\phi}_{j} } \sum\limits_{n=1}^N{ \log c_{j}(\hat{F}_{1}(z_{1n}),\ldots,\hat{F}_{|\mathcal G_k|}(z_{|\mathcal G_k|n})|\bs{\phi}_{j}) }.
\end{equation}

With the estimated parameter(s), the best copula $c^*$ is given as
\begin{equation}
c^* = \arg \min_{c_{j} \in \mathcal{C}} \text{AIC}_{j}.
\label{eq:bestccc}
\end{equation}

The $\text{AIC}$ value  is given as
\begin{equation}
\label{eq:aiccc}
\text{AIC}_{j} = - \sum_{n=1}^N{ \log{c_{j}( \hat{F}_{1}(z_{1n}),\ldots, \hat{F}_{|\mathcal G_k|}(z_{|\mathcal G_k|n}) | \widehat{\bs{\phi}}_{j})}} + q(c_{j}),
\end{equation}
where $q(c_{j})$ is the number of parameters in the $j$th copula model.

\subsection{Cluster Based Consensus Scheme}
After all the clusters obtain their initial estimates, these estimates are shared via linear inter-cluster collaboration to reach a consensus. We employ the average consensus algorithm \cite{olfati2007consensus}. 
Assume that the collaboration among clusters is represented by a fixed topology matrix $\mathbf A$ with binary entries, namely, $A_{ij} \in \{ 0, 1\}, i, j \in [K]$. Here, $A_{ij} = 1$ means that there is a communication link from the $i$th cluster to the $j$th cluster; otherwise, $A_{ij} = 0$. At iteration $n+1$, each cluster $\mathcal G_i, i \in [K]$ updates its estimate $\hat{\theta}_{\mathcal G_i}(n+1)$ as follows \cite{olfati2007consensus}:
\begin{equation}
\hat{\theta}_{\mathcal G_i}(n+1) = \hat{\theta}_{\mathcal G_i}(n) - \beta \sum_{j \in N_{\mathcal G_i}} A_{i j} \left( \hat{\theta}_{\mathcal G_j}(n) -  \hat{\theta}_{\mathcal G_i}(n) \right),
\end{equation}
where $0 < \beta < 1/ \Delta$, $\Delta$ is the maximum degree of the network and $N_{\mathcal G_i}$ is the neighborhood cluster set of $\mathcal G_i$.

It has been shown in \cite[Theorem 2]{olfati2007consensus} that if the graph is strongly connected and balanced,  $\hat{\theta} = \frac{\sum_{i}\hat{\theta}_{\mathcal G_i}(0)}{K}$ asymptotically.

\begin{theorem}
\label{thm:var}
The standard deviation of the parameter estimate obtained by the average consensus scheme is upper bounded by the average standard deviation of all the clusters' estimates, i.e.,
\begin{equation}
\sqrt{\text{var}(\hat{\theta})} \leq \frac{1}{K} \sum_{k=1}^{K}\sqrt{\text{var}(\hat{\theta}_k) }
\end{equation}
where $k, k \in [K]$ denotes the cluster index and $\text{var}(\cdot)$ represent the variance of a random variable.
\end{theorem}
\textbf{Proof}: See Appendix \ref{appendix:thmvar}. \hfill $\blacksquare$

\begin{remark}
The average consensus based inter-cluster collaboration helps in mitigating the effect of estimation bias resulting from the individual cluster estimates.
\end{remark}

Since the sensor network is large, the number of sensors in each cluster is also potentially large. As mentioned in Section \ref{sec:i}, some sensors may provide redundant information. Allowing all the sensors in the cluster to exchange their information may result in a large transmission cost. Therefore, selecting sensors with maximum information and minimum redundancy is crucial.  
In the following section, we propose a mutual information based sensor selection scheme, and only the selected sensors need to exchange their information within a cluster. 

\begin{remark}
In practice, to extend the network lifetime, one may design sleep scheduling schemes for sensors which provide redundant data \cite{deng2005scheduling, deng2005balanced}. Also, to balance battery usage for inter-cluster communication, one may rotate sensors that are responsible for inter-cluster collaboration in a small region near the edge of the cluster.
\end{remark}

\section{Sensor Selection Based Two-Step Dependence Driven Collaborative Distributed Estimation}
\label{sec:sensorselection}
In this section, we present the details of the sensor selection scheme for our two-step collaborative estimation scheme.
For each cluster $\mathcal G_k, k \in [K]$, prior to estimation via intra-cluster collaboration, a mutual information based methodology is employed to select sensors with maximum information and minimum redundancy.

Before we proceed, we recall that the mutual information of two random variables $x$ and $y$, denoted by $I(x; y)$, is given as
\begin{equation}
I(x; y) = \int f(x, y) \text{log} \left (\frac{f(x, y)}{f(x) f(y)} \right) dx dy,
\end{equation} 
where $f(x, y)$ is the joint PDF of variables $x$ and $y$. $f(x)$ and $f(y)$ are the marginal PDFs.

The optimal sensor selection strategy is often based on maximal relevance and minimal redundancy with respect to the target parameter $\theta$ on the entire sensor set \cite{peng2005feature}, and this strategy is referred as \textit{maximal-relevancy-minimal-redundancy} (mRMR) in \cite{peng2005feature}.  Suppose that we aim to select $m$ sensors from the set of all the sensors in the network $\mathcal S$  
with the corresponding observation set $\mathbf z_{\mathcal S} = \{ \mathbf z_1, \ldots, \mathbf z_{|\mathcal S|} \}$. In terms of mutual information, the mRMR solution is obtained by solving the following problem
\begin{equation}
\label{eq:oneMI}
\underset{\mathbf s_m \in \mathcal S}{\max} \, \left [\frac{1}{|\mathbf s_m|} \underset{\mathbf z_i \in \mathbf z_{\mathbf s_m}}{\sum}I(\mathbf z_i; \theta) - \frac{1}{|\mathbf s_m|^2}\underset{\mathbf z_i, \mathbf z_j \in \mathbf z_{\mathbf s_m}}{\sum}I(\mathbf z_i; \mathbf z_j) \right],
\end{equation} 
where $\mathbf s_m$ is the set of the selected sensors with cardinality $|\mathbf s_m| = m$ and $\mathbf z_{\mathbf s_m} = \{ \mathbf z_1, \ldots, \mathbf z_{|\mathbf s_m|} \}$ is the sensor observation set of  $\mathbf s_m$, where $\mathbf z_{\mathbf s_m} \in \mathbf z_{\mathcal S}$.  

Note that the computational complexity of the mRMR problem is $O(|\mathcal S|^m)$. A more efficient first-order incremental search method was proposed to find the near-optimal solutions of problem in Equation \eqref{eq:oneMI} in \cite{peng2005feature}.  It is given as:
\begin{equation}
\label{eq:increO}
\underset{\mathbf z_j \in \mathbf z_{\mathcal S} \setminus \mathbf z_{\mathbf s_{m-1}} }{\mymax} \left [ I(\mathbf z_j; \theta) - \frac{1}{m-1} \underset{\mathbf z_i \in \mathbf z_{\mathbf s_{m-1}}}{\sum} I (\mathbf z_j ; \mathbf z_i)\right],
\end{equation}
where $\mathbf s_{m-1}$ is the selected sensor set with $m-1$ sensors, and $\mathbf z_{\mathcal S} \setminus \mathbf z_{\mathbf s_{m-1}}$ denotes that we exclude the sensor observations from the sensors in set $ \mathbf s_{m-1}$ from $\mathbf z_{\mathcal S}$.

The computational complexity of the incremental search method in Equation \eqref{eq:increO} is $O(m * |\mathcal S|)$. To further reduce the computational complexity,  in the following, we propose a cluster-based incremental search methodology, where the sensor selection is performed cluster-by-cluster independently.

Note that $\mathcal S = \{ \mathcal G_1 \cup \mathcal G_2 \ldots \cup \mathcal G_K\}$, where $\mathcal G_i \cap \mathcal G_j = \emptyset, i, j \in [K]$. Instead of searching over the entire sensor set, we select $m_k$ sensors from cluster $\mathcal G_k, k \in [K]$. Note that $\sum_{k=1}^{K} m_k = m$.

For each cluster $\mathcal G_k$, suppose that we already have $\mathbf s_{m_{k}-1}$, the sensor set with $m_{k}-1$ sensors.  The incremental selection scheme solves the following problem:
\begin{equation}
\label{eq:incre}
\underset{\mathbf z_j \in \mathbf z_{\mathcal G_k} \setminus \mathbf z_{\mathbf s_{m_{k}-1}} }{\mymax} \left [ I(\mathbf z_j; \theta) - \frac{1}{m_{k}-1} \underset{\mathbf z_i \in \mathbf z_{\mathbf s_{m_{k}-1}}}{\sum} I (\mathbf z_j ; \mathbf z_i)\right],
\end{equation}
where $\mathbf z_{\mathcal G_{k}} \setminus \mathbf z_{\mathbf s_{m_{k}-1}}$ denotes that we exclude the sensor observations in set $ \mathbf s_{m_{k}-1}$ from $\mathbf z_{\mathcal G_{k}}$.

\begin{remark}
The selection scheme given in Equation \eqref{eq:increO} is referred to as the global sensor selection scheme. Also, the selection scheme given in Equation \eqref{eq:incre} is referred to as the cluster-based sensor selection scheme.
\end{remark}

\begin{theorem}
\label{theorem:group}
Using a suitably designed threshold $d_{th}$ that makes inter-cluster sensors nearly independent, the cluster-based sensor selection method is equivalent to the global sensor selection method with probability at least $1 - \epsilon$, where $\epsilon$ is a small constant.
\end{theorem}
\textbf{Proof}: See Appendix \ref{appendix:thmgroup}. \hfill $\blacksquare$

A natural question is how to determine the optimal number of sensors $m_k$ for cluster $\mathcal G_k, k \in [K]$. In an energy constrained network with battery-limited sensors, each sensor's energy is finite and a communication cost is incurred when it transmits observations to collaborating sensors. Therefore, the number of sensors that can be selected in each cluster is limited due to finite energy budgets. Let $r$ be the average number of requests initiated by each sensor in the network per unit time interval. Then, for the selected sensors in cluster $\mathcal G_k$, the number of requests that have to be responded to within a unit time interval is $r(m_k - 1)$. Moreover, we assume that the energy cost for a single transmission is $E_t$. The average energy consumption per unit time interval for each selected sensor in cluster $\mathcal G_k$ is $\mathbb E[s] = r(m_k - 1) E_t, s \in \mathbf s_{m_k}$, which increases as the size of the selected sensor set $\mathbf s_{m_k}$ increases. Let the energy consumption of cluster $\mathcal G_k$ be the average energy consumption per sensor in $\mathbf s_{m_k}$, denoted by $\mathbb E[\mathbf s_{m_k}]$. Thus, in terms of energy efficiency, a smaller sensor set is preferred. In order to guarantee adequate sensors lifetimes, we enforce the energy consumption constraint as follows:
\begin{equation}
\label{eq:E_s}
\mathbb E[\mathbf s_{m_k}] = r(m_k - 1) E_t \leq \alpha_k, k \in [K],
\end{equation}
where $\alpha_k > 0$ is the pre-specified constraint for cluster $k$. It should be noted that $m_k=1$ always satisfies the energy consumption constraint \eqref{eq:E_s}, i.e., at least one sensor is selected from each cluster.
Therefore, the energy constrained selection scheme for each cluster $\mathcal G_k, k \in [K]$ is stated as
\begin{align}
\label{eq:energyp}
&\underset{\mathbf s_{m_k} \in \mathcal G_k}{\max} \quad \quad \,\, \frac{1}{m_k} \underset{\mathbf z_i \in \mathbf z_{\mathbf s_{m_k}}}{\sum} I (\mathbf z_i; \theta) - \frac{1}{m_k^2} \underset{\mathbf z_i, \mathbf z_j \in \mathbf z_{\mathbf s_{m_k}}}{\sum} I(\mathbf z_i, \mathbf z_j),  \\
&\text{subject to} \quad \mathbb E[\mathbf s_{m_k}] \leq \alpha_k. \nonumber
\end{align}

The problem in Equation \eqref{eq:energyp} can be solved using the incremental search method in Equation \eqref{eq:incre} while satisfying the energy constraint. 

\section{Numerical Results}
\label{sec:nr}
In this section, we demonstrate the efficacy of our proposed two-step cluster-based collaborative distributed estimation methodologies through numerical examples. We consider a wireless sensor network with $L = 13$ sensors deployed in a $[0, 1.5] \times [0, 1.5]$ square area of interest.  Let $(x_0, y_0)$ be the target location coordinates and $\theta$ be the intensity of the target signal to be estimated.  We assume a Gaussian prior $\mathcal N(\theta_p, \sigma^2_p)$ on $\theta$. 
Sensor $l, l = 1, \ldots, L$ is located at $(x_{l}, y_{l})$. The received measurements at the $l$th sensor are modeled as
\begin{equation}
z_{li} = \theta + w_{li}, i = 1, \ldots, N,
\end{equation}
where $w_{li}$ is the measurement noise which is assumed to be Gaussian distributed with mean $0$ and variance $\sigma_l^2$ and $N$ is the number of observations. Here, we assume that the variance of the measurement noise at each sensor is inversely scaled by the distance between the sensor and the signal source, i.e., $\sigma_l^2 = \frac{\sigma_0^2}{\sqrt{(x_l - x_0)^2 + (y_l - y_0)^2}}$. 
Note that $\sigma_0^2$ is introduced here for the ease of characterizing signal to noise ratio (SNR) at different sensors. We define our SNR as 
\begin{equation}
\text{SNR} = \frac{\mathbb E[\theta^2]}{\sigma_0^2}.
\end{equation}

We assume that the measurement noise $w_{li}$ and $\theta$ are independent of each other. Moreover, we assume that the measurement noises are i.i.d. across time and can be spatially dependent at some sensors.  Without loss of generality, we assume that we have three clusters and the underlying spatial dependence among sensors is generated cluster by cluster using multivariate Clayton copula functions. The pair-wise sensor dissimilarities are estimated based on Kendall's $\tau$. We set $r E_t = 1$. Therefore, according to Equation \eqref{eq:E_s}, the maximum number of sensors that can be selected in cluster $k, k \in [K]$ is $m_k = \alpha_k$. Also, without loss of generality, we assume that $\alpha_1 = \alpha_2 = \ldots, \alpha_K$. Therefore, $m_1 = m_2 = \ldots, m_K$. The total number of sensors that are selected is $m = \sum_{k=1}^{K} m_k$.

We use average mean squared error (MSE) to characterize the estimation performance. For the clustering process, we use the average clustering accuracy to measure the clustering performance. The clustering accuracy is defined as $\frac{\text{Number of correctly clustered sensors}}{\text{Total number of sensors}}$. All the results are obtained using $500$ Monte Carlo trials.
 
To exhibit the performance improvement by applying our proposed two-step cluster-based collaborative distributed estimation methodologies, we also evaluate the corresponding estimation performance under independence assumption that ignores dependence among sensor observations. Moreover, we compare our proposed estimation methodologies with the cluster-based MAP method given in Equation \eqref{eq:joint2}, where the copula-based approach as well as the product-based approach (under independence assumption) can be used to model the conditional joint PDFs.
For clarity, we summarize the eight empirically studied cases as follows.
\begin{itemize}
\item Cluster-based consensus with sensor selection using copula based method as well as under independence assumption
\item Cluster-based consensus without sensor selection using copula based method as well as under independence assumption
\item Cluster-based MAP with sensor selection using copula based method as well as under independence assumption
\item Cluster-based MAP without sensor selection using copula based method as well as under independence assumption
\end{itemize}

\begin{figure}
\centering
\includegraphics[height=2.4in,width=!]{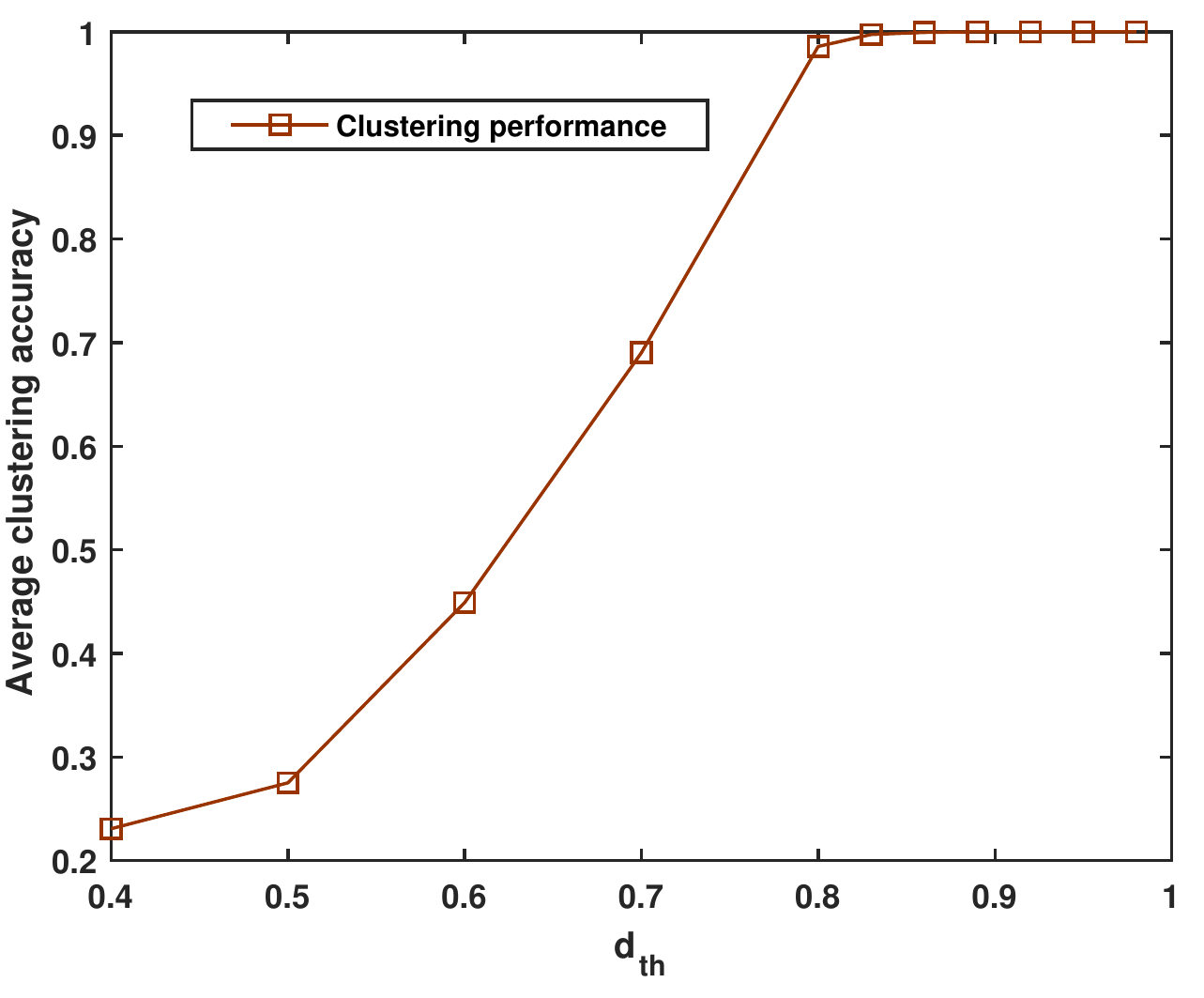}
\caption{Average clustering accuracy as a function of threshold $d_{th}$.}
\label{fig:d_th_error}
\end{figure}

\begin{figure}
\centering
\includegraphics[height=2.4in,width=!]{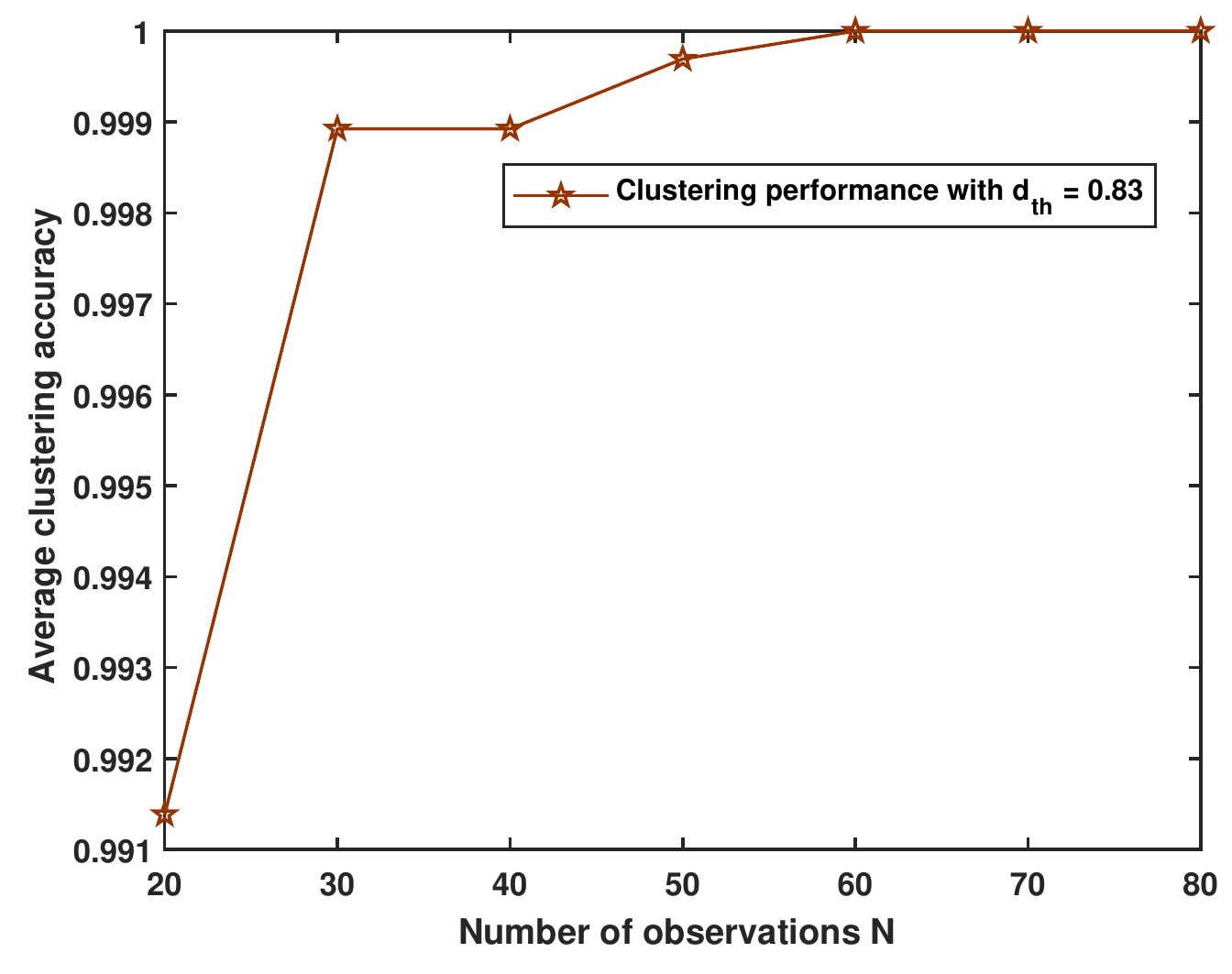}
\caption{Average clustering accuracy as a function of number of observations $N$ with $d_{th} = 0.83$.}
\label{fig:number_observations}
\end{figure}

In Fig.\,\ref{fig:d_th_error}, we present the average clustering accuracy as a function of the threshold $d_{th}$ at $\text{SNR} = 2.0$ dB and $N = 70$. We can see that the choice of $d_{th}$ has a significant impact on the performance of Algorithm \ref{algo:cluster}. The optimal value of $d_{th}$ depends on the given data, namely, $d_L$ and $d_H$. Moreover, as we can see, a larger $d_{th}$ results in a better clustering performance. 

In Fig.\,\ref{fig:number_observations}, we present the average clustering accuracy as a function of the number of observations $N$ with $d_{th} = 0.83$ at $\text{SNR} = 2.0$ dB. As we can see, by choosing appropriate $d_{th}$ and $N$, we can achieve perfect clustering performance. In the following, our estimation results are obtained with $d_{th} = 0.83$ and $N = 70$ unless otherwise specified.

\begin{figure}
\centering
\includegraphics[height=2.4in,width=!]{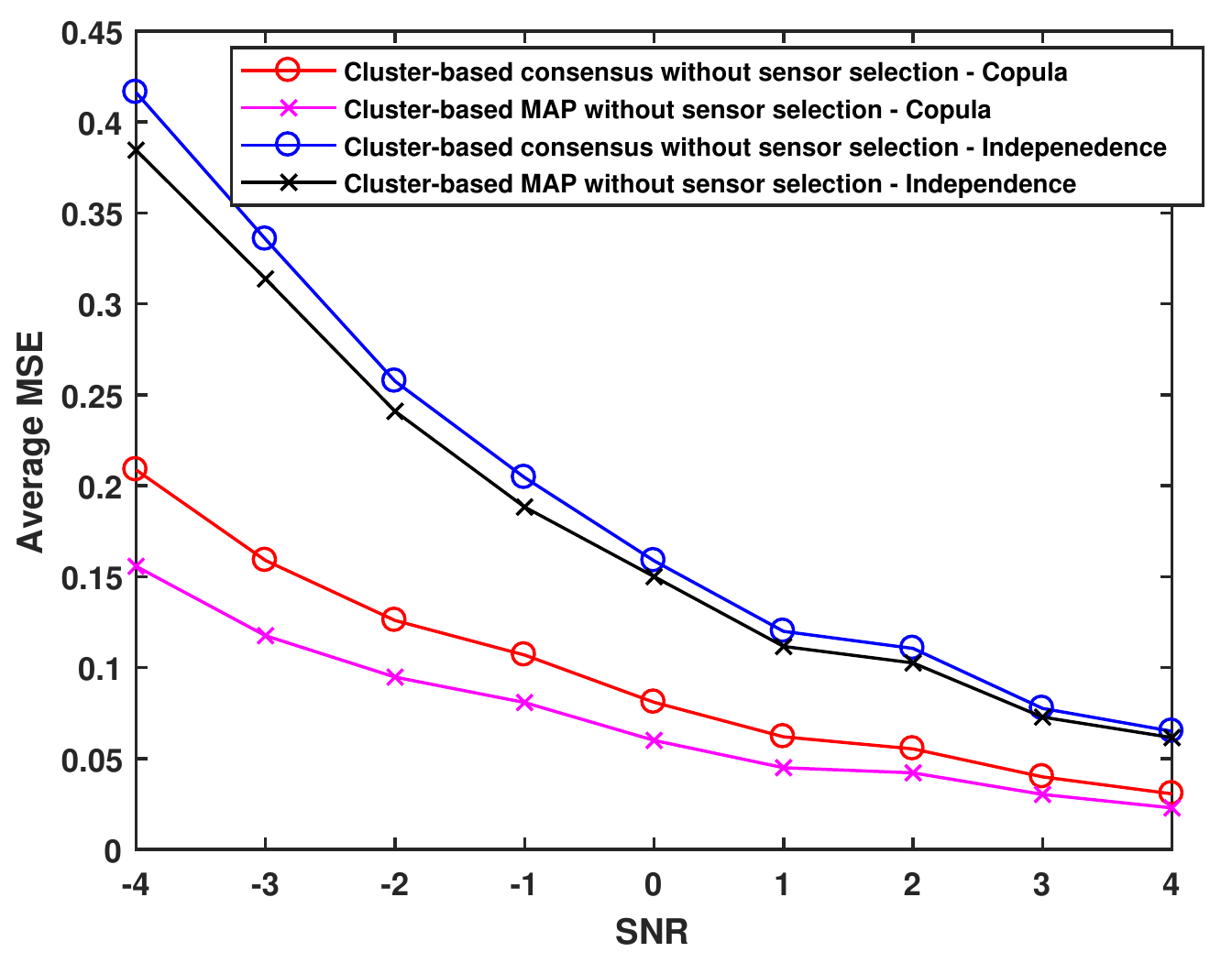}
\caption{Average MSE as a function of SNR without sensor selection.}
\label{fig:mse_without}
\end{figure}

\begin{figure}
\centering
\includegraphics[height=2.4in,width=!]{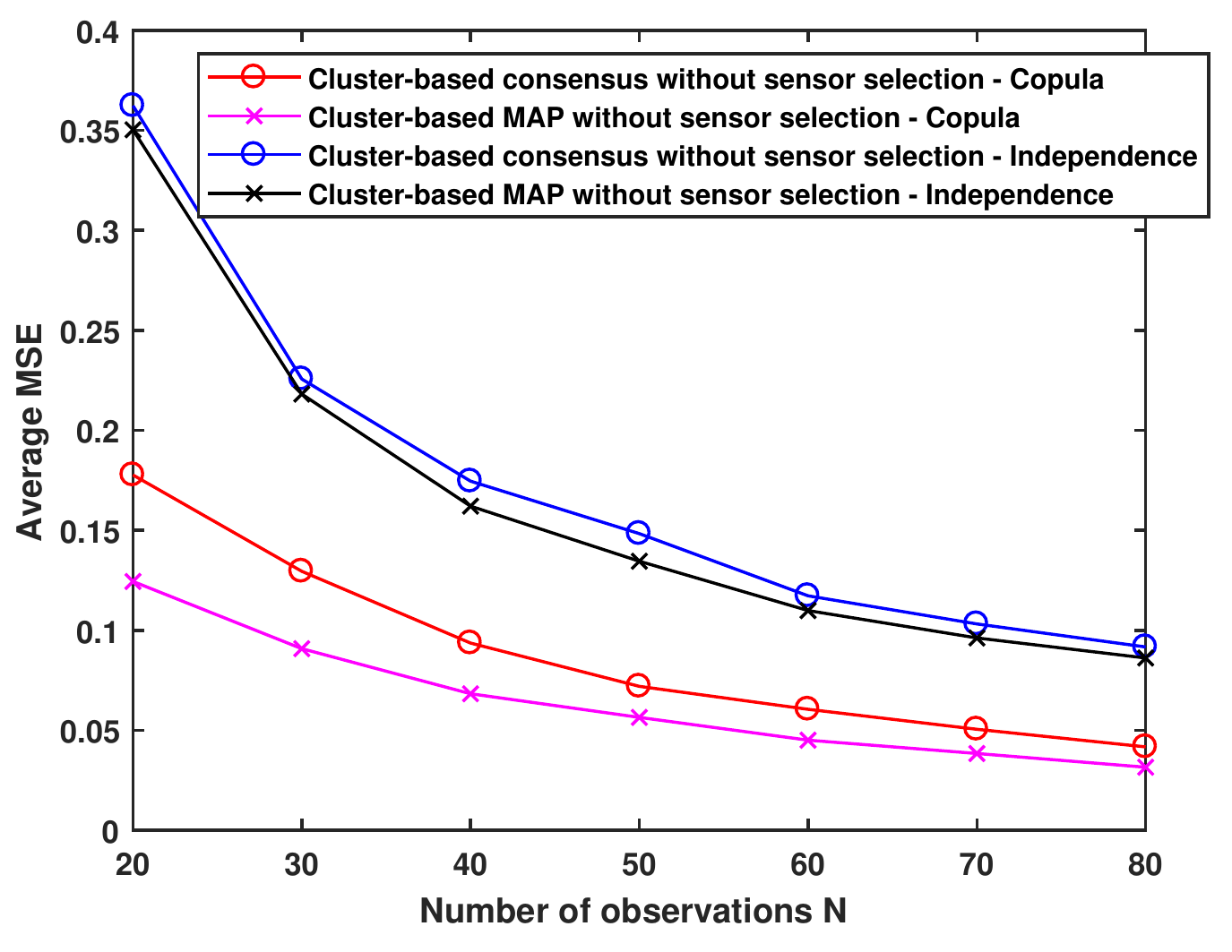}
\caption{Average MSE as a function of the number of observations $N$ without sensor selection.}
\label{fig:mse_withoutN}
\end{figure}

\begin{figure}
\centering
\includegraphics[height=2.4in,width=!]{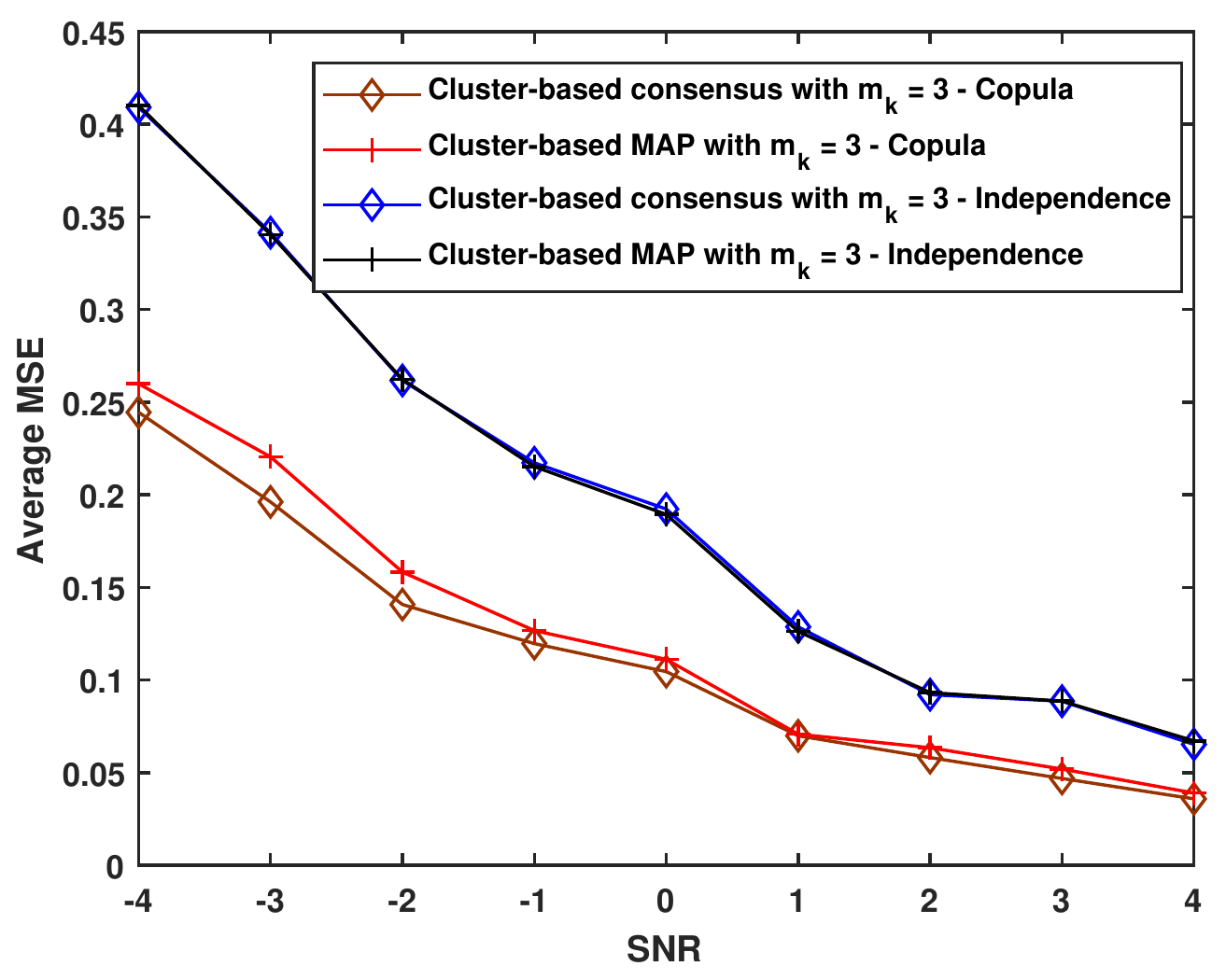}
\caption{Average MSE as a function of SNR with cluster-based sensor selection and $m_k = 3$.}
\label{fig:mse_with3}
\end{figure}

\begin{figure}
\centering
\includegraphics[height=2.4in,width=!]{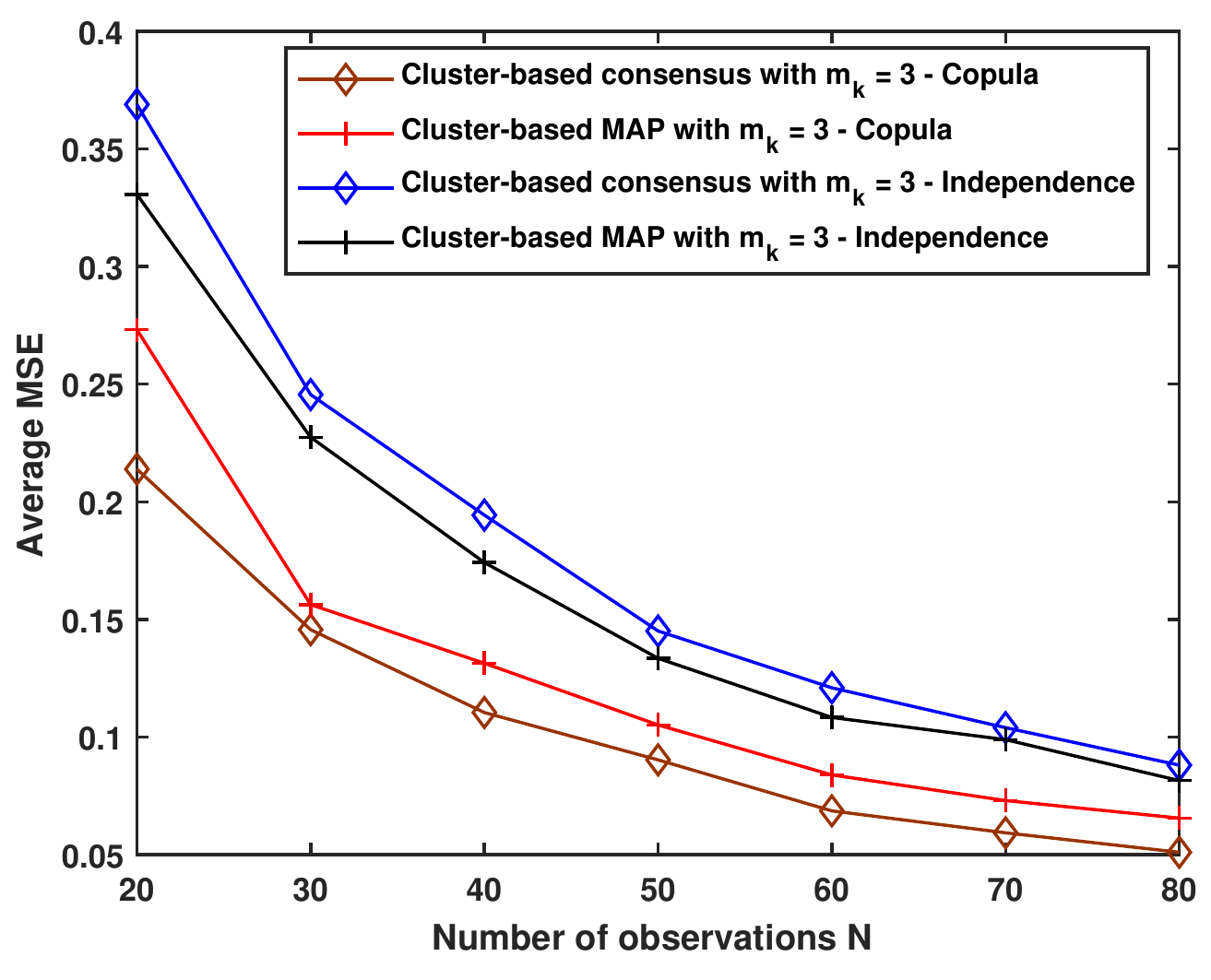}
\caption{Average MSE as a function of the number of observations $N$ with cluster-based sensor selection and $m_k = 3$.}
\label{fig:mse_with3N}
\end{figure}

\begin{figure}
\centering
\includegraphics[height=2.4in,width=!]{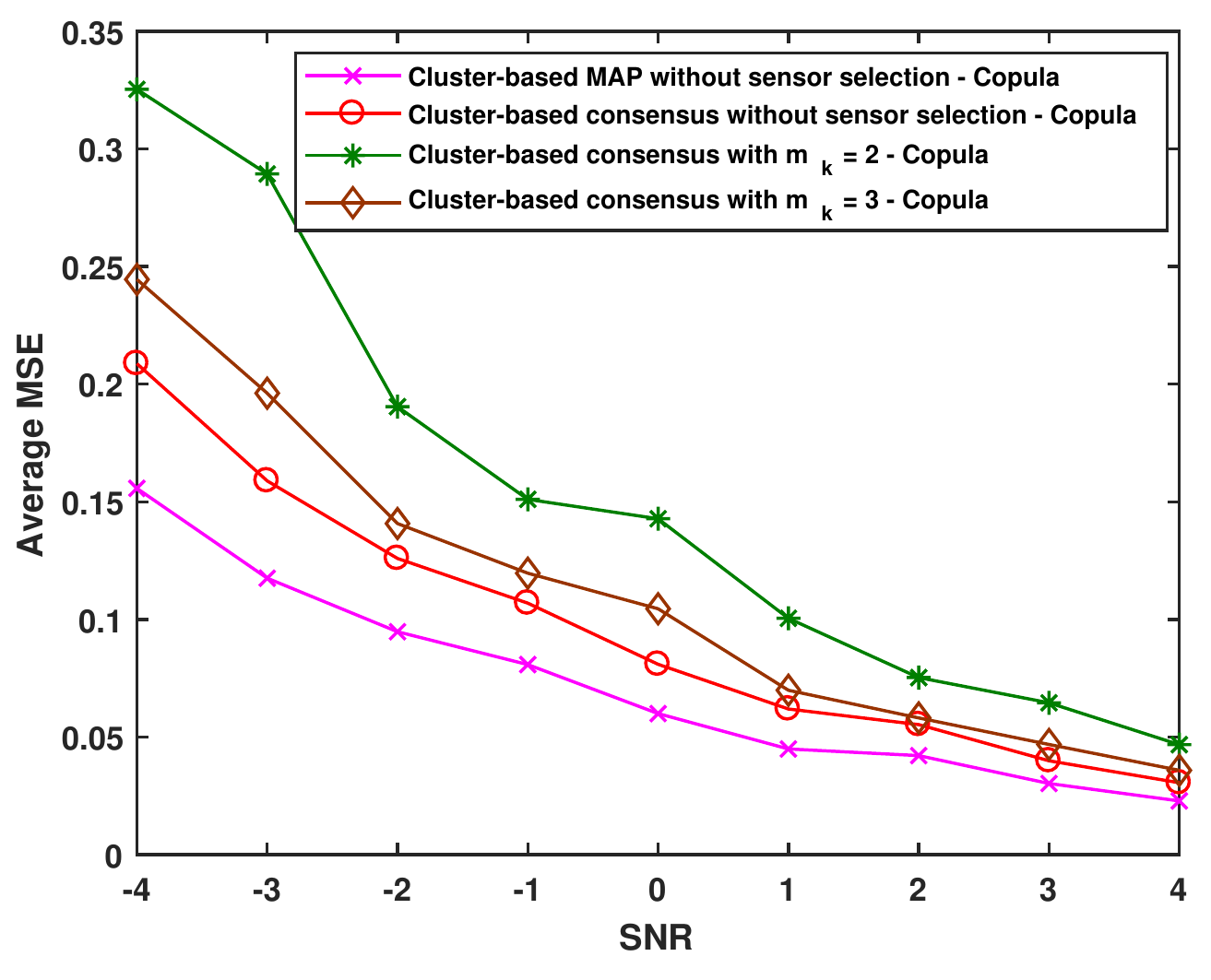}
\caption{Average MSE as a function of SNR for different  schemes without sensor selection.}
\label{fig:mse_compare}
\end{figure}

\begin{figure}
\centering
\includegraphics[height=2.4in,width=!]{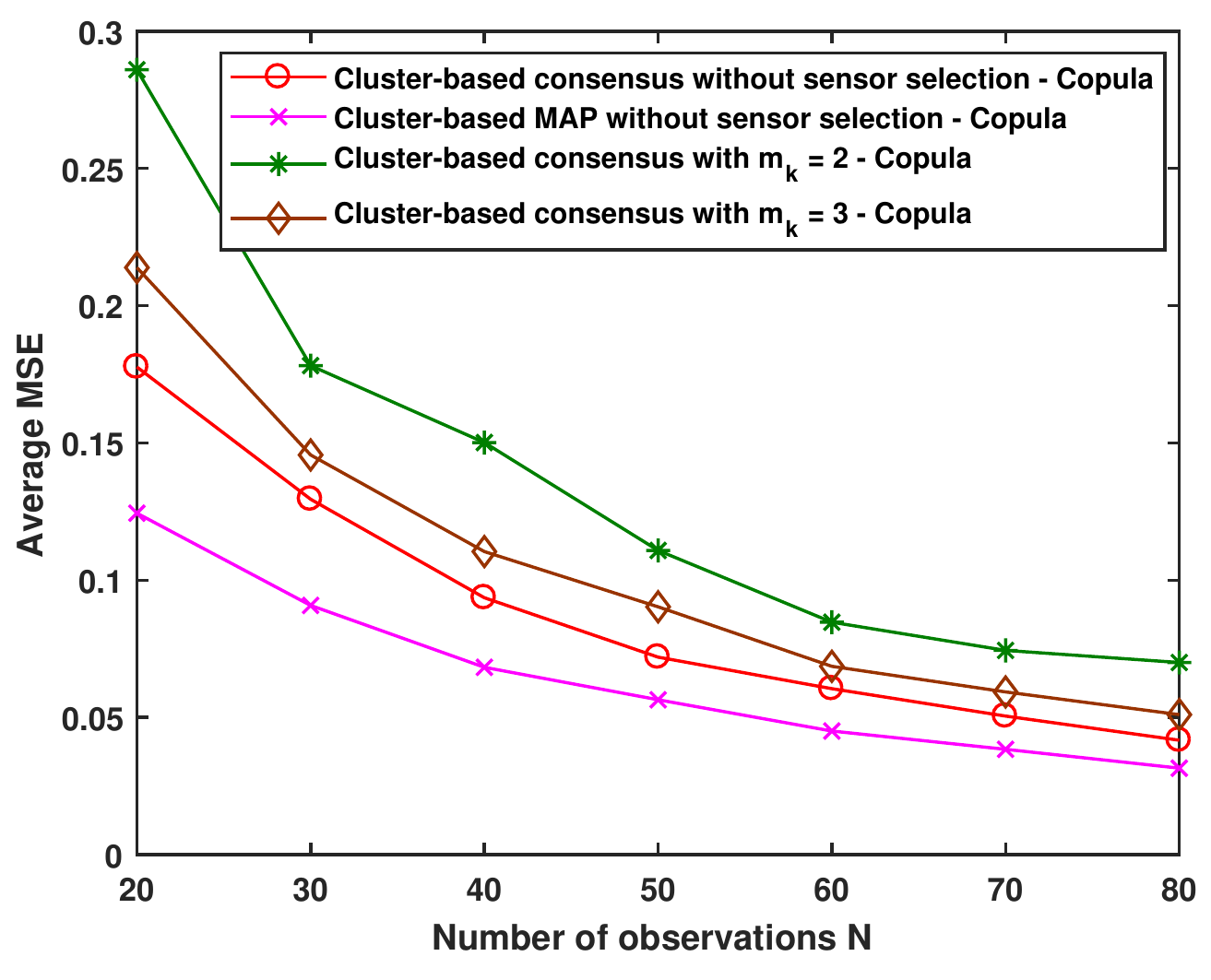}
\caption{Average MSE as a function of the number of observations $N$ for different  schemes without sensor selection.}
\label{fig:mse_compareN}
\end{figure}

\begin{figure}
\centering
\includegraphics[height=2.4in,width=!]{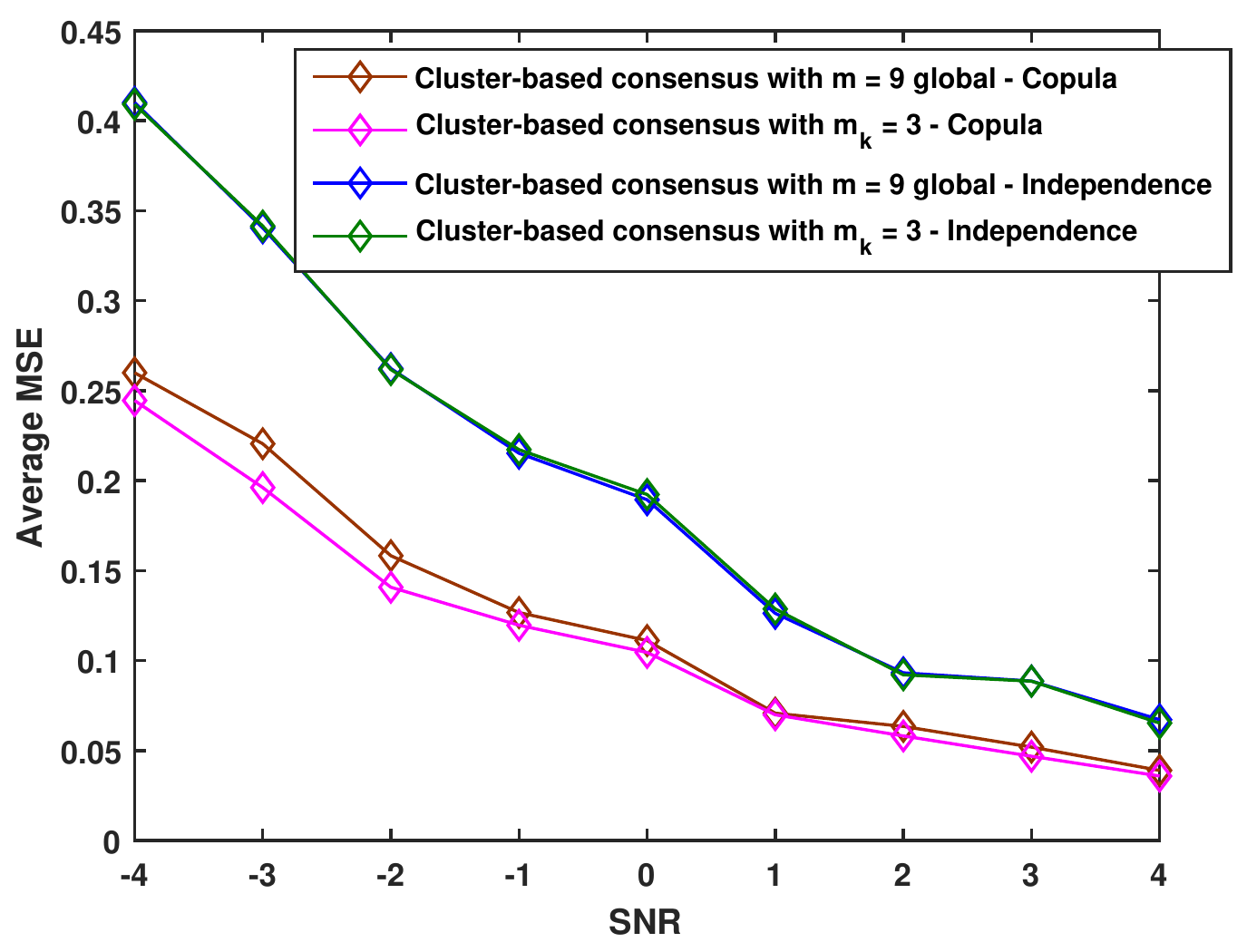}
\caption{Average MSE as a function of SNR for the cluster-based sensor selection scheme and the global sensor selection scheme.}
\label{fig:mse}
\end{figure}

\begin{figure}
\centering
\includegraphics[height=2.4in,width=!]{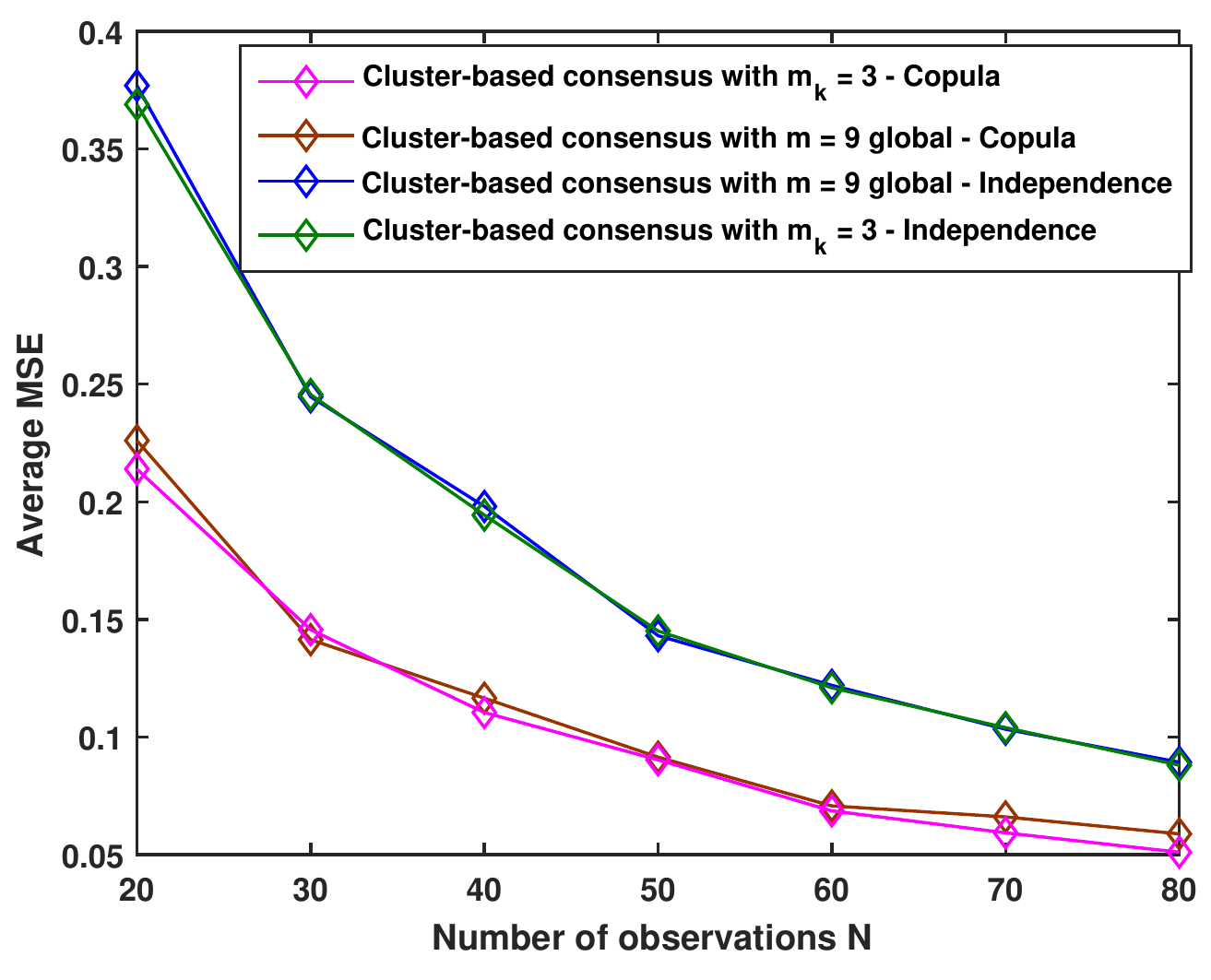}
\caption{Average MSE as a function of the number of observations $N$ for the cluster-based sensor selection scheme and the global sensor selection scheme.}
\label{fig:mse_N}
\end{figure}


In Fig.\,\ref{fig:mse_without} and Fig.\,\ref{fig:mse_withoutN}, we present the average MSE as a function of SNR and the number of observations $N$, respectively, and compare the performance of schemes without sensor selection. The schemes that are evaluated are: Cluster-based consensus without sensor selection using copula based scheme, Cluster-based MAP without sensor selection using copula based scheme, Cluster-based consensus without sensor selection under independence assumption and Cluster-based MAP without sensor selection under independence assumption. We can see that as $N$ as well as $\text{SNR}$ increases, the average MSE decreases.  Also, the schemes using copula based estimation methodologies perform significantly better than the schemes that assume independence among sensor observations. Moreover, as we can see, for the independent cases, the cluster-based consensus scheme performs pretty close to the cluster-based MAP scheme while for the copula cases, the cluster-based consensus scheme performs close to the corresponding cluster-based MAP scheme at the SNR values greater than $0$ dB in Fig.\,\ref{fig:mse_without} and the number of observations larger than $50$ in Fig.\,\ref{fig:mse_withoutN}. Note that with extremely low SNR values or very small number of observations, the estimation performance difference between the copula incorporated cluster-based consensus scheme and the copula incorporated cluster-based MAP scheme is large. This is because for the cluster-based consensus scheme, the estimate obtained from each cluster is relatively poor for  extremely low SNR values or with very small number of observations while for the cluster-based MAP scheme, it models the conditional joint PDF and captures more information.

In Fig.\,\ref{fig:mse_with3} and Fig.\,\ref{fig:mse_with3N}, we present the average MSE as a function of SNR and the number of observations $N$, respectively, by comparing schemes with cluster-based sensor selection. The schemes that are evaluated are: Cluster-based consensus with $m_k = 3$ using copula based scheme, Cluster-based MAP with $m_k = 3$ using copula based scheme, Cluster-based consensus with $m_k = 3$ under independence assumption and Cluster-based MAP with $m_k = 3$ under independence assumption. 
As we can see,  the schemes using copula based estimation methodologies perform significantly better than the schemes assuming independence among sensor observations. 
Note that with sensor selection, the cluster-based consensus scheme using copula incorporated estimation methodology performs better than the corresponding cluster-based MAP scheme. This is because our proposed sensor selection scheme aims to select sensors with maximum relevance and minimum redundancy (namely, most independent sensors) regarding the parameter of interest. With most independent selected sensors, part of the dependence information for each cluster is lost.  For the cluster-based MAP scheme, the product approach is used to combine the  conditional joint PDFs corresponding to each cluster whereas for the cluster-based consensus scheme, consensus is used and the estimates obtained from each cluster are linearly combined where the linear dependence is imposed inherently resulting in better performance.

In Fig.\,\ref{fig:mse_compare} and Fig.\,\ref{fig:mse_compareN}, we present the average MSE as a function of SNR and the number of observations $N$, respectively, for copula incorporated schemes with cluster-based sensor selection and the copula incorporated schemes without sensor selection. The schemes that are evaluated are: Cluster-based consensus without sensor selection using copula based scheme, Cluster-based MAP without sensor selection using copula based scheme, Cluster-based consensus with $m_k = 2$ using copula based scheme and Cluster-based consensus with $m_k = 3$ using copula based scheme. As we can see that, selecting $m_k = 3$ sensors in each cluster results in better estimation performance compared to selecting $m_k = 2$ sensors in each cluster. Moreover, in Fig.\,\ref{fig:mse_compare}, our proposed cluster-based consensus approach by selecting $m_k = 3$ sensors in each cluster performs very close to the corresponding scheme without sensor selection for SNR from $1$ dB to $4$ dB. Also, we have similar performance in Fig.\,\ref{fig:mse_compareN} when the number of observations is larger than or equal to $60$. For the SNR value smaller than $1$ dB and the number of observations smaller than $60$, the performance difference between the cluster-based consensus scheme by selecting $m_k = 3$ sensors and the cluster-based consensus scheme without sensor selection is large. This is due to the fact that with low SNR values or small number of observations, the estimate obtained from each cluster is relatively poor. However, for the corresponding scheme without sensor selection, it includes more sensors and contains more information.

In Fig.\,\ref{fig:mse} and Fig.\,\ref{fig:mse_N}, we present the average MSE as a function of SNR and the number of observations $N$, respectively, for the cluster-based sensor selection scheme and the global sensor selection scheme (see Equation \eqref{eq:increO}). We evaluate the following schemes: Cluster-based consensus scheme using global sensor selection with $m = 9$ and copula based approach, Cluster-based consensus scheme using cluster-based selection scheme with $m_k = 3$ and copula based approach, Cluster-based consensus scheme using global sensor selection with $m = 9$ under independence assumption and Cluster-based consensus scheme using cluster-based selection scheme with $m_k = 3$ under independence assumption. Note that for fair comparison of the cluster-based sensor selection scheme and the global sensor selection scheme, we set $m = 9$ as the total number of sensors that are selected since $m_k = 3$ sensors are selected from each cluster and the estimated number of clusters is $\hat{K} = 3$.
 As we can see, the cluster-based sensor selection scheme and global sensor selection scheme perform equally well.

\section{Conclusion}
\label{sec:conclusion}
In this paper, a two-step cluster-based collaborative distributed estimation scheme was presented, where in the first step, sensors first form dependence driven clusters, and then perform copula-based MAP estimation via intra-cluster collaboration; in the second step, the estimates generated in the first step are shared via inter-cluster collaboration until an average consensus is reached. We proposed a merge based $K$-medoid dependence driven clustering algorithm. 
We further proposed a cluster-based sensor selection incorporated collaborative distributed estimation scheme. More specifically, prior to estimation, each cluster employs a mutual information based sensor selection scheme and selects  sensors with maximum relevance and minimum redundancy with respect to the target parameter.  Also, the proposed cluster-based sensor selection scheme was shown to be equivalent to the global based selection scheme with high probability, and was computationally more efficient.
Numerical results demonstrated the efficiency of our proposed methods compared to the estimation schemes under independence assumption. 

In the future, one can consider a sparsity imposed copula-based scheme for the problem of distributed estimation in large scale sensor network.

\appendices
%

\section{Proof of Theorem \ref{thm:var}}
\label{appendix:thmvar}
\begin{small}
\begin{align*}
\text{var}\left (\hat{\theta} \right) &= \text{var} \left (\frac{1}{K} \sum_{k=1}^{K} \hat{\theta}_k \right), \\ \nonumber
&= \frac{1}{K^2} \left [\sum_{k=1}^{K} \text{var}\left (\hat{\theta}_k \right) + \sum_{k=1}^{K} \sum_{\tilde{k} \neq k = 1}^{K}\text{cov}\left (\hat{\theta}_k, \hat{\theta}_{\tilde k} \right) \right ], \\ \nonumber
&\overset{(b)}{\leq} \frac{1}{K^2} \left [ \sum_{k=1}^{K} \text{var}\left (\hat{\theta}_k \right) +  \sum_{k=1}^{K} \sum_{\tilde{k} \neq k = 1}^{K} \sqrt{\text{var}\left ( \hat{\theta}_k \right) \text{var}\left (\hat{\theta}_{\tilde k} \right) } \right ], \\ \nonumber
&= \left ( \frac{1}{K} \sum_{k=1}^{K} \sqrt{\text{var}\left (\hat{\theta}_k \right)} \right)^2,
\end{align*}
where $(b)$ is obtained using $\text{cov}\left (\hat{\theta}_k, \hat{\theta}_{\tilde k} \right) / \sqrt{\text{var}\left ( \hat{\theta}_k \right) \text{var}\left (\hat{\theta}_{\tilde k} \right)} \leq 1$. Thus, we obtain that 
\begin{equation*}
\sqrt{\text{var}\left (\hat{\theta} \right)} \leq \frac{1}{K}\sum_{k=1}^{K}\sqrt{\text{var}\left (\hat{\theta}_k \right)}.
\end{equation*}
\end{small}
\hfill $\blacksquare$

\section{Proof of Theorem \ref{theorem:group}}
\label{appendix:thmgroup}
Suppose that we already have the set $\mathbf s$ which consists of selected sensors using the global incremental sensor selection scheme in \eqref{eq:increO}.  We can trace back these selected sensors in the set $\mathbf s$ to clusters.  Without loss of generality, we assume that the sensors in the set $\mathbf s$ belong to clusters $\mathcal G_1, \ldots, \mathcal G_i, i < K$, and we decompose the set $\mathbf s$ into $\mathbf s_1, \ldots, \mathbf s_i$ with $s_r, r = 1, \ldots, i$ denoting the subset of sensors that belongs to cluster $\mathcal G_r$.  

Assume that we have a candidate data sequence $\mathbf z_{\tilde j} \in \mathbf z_{\mathcal S} \setminus \mathbf z_{\mathbf s}$, where $\mathbf z_{\mathcal S}$ is the set of data sequences obtained from all the sensors in the network. 
Therefore, the global incremental selection problem becomes:
\begin{equation}
\label{eq:glo}
\underset{\mathbf z_{\tilde j} \in \mathbf z_{\mathcal S} \setminus \mathbf z_{\mathbf s} }{\mymax}  I(\mathbf z_{\tilde j}; \theta) - \frac{1}{|\mathbf s|} \underset{\mathbf z_{t} \in \mathbf z_{\mathbf s}}{\sum} I (\mathbf z_{\tilde j}; \mathbf z_{t}).
\end{equation}

Note that there are two cases for the assignment of the sequence $\mathbf z_{\tilde j}$ . The first case is that $\mathbf z_{\tilde j}$ belongs to one of the clusters in set $[\mathcal G_1, \ldots, \mathcal G_i]$. The second case is that $\mathbf z_{\tilde j}$ belongs to one of the clusters in set $[\mathcal G_{i+1}, \ldots, \mathcal G_K]$. 

For the first case, without loss of generality, we assume that $\mathbf z_{\tilde j}$ belongs to cluster $\mathcal G_j$. Also, we further suppose that set $\mathbf s_{{j}} \in \mathbf s$ contains the selected sensors from cluster $\mathcal G_{j}, j \leq i$. Thus, the problem in Equation \eqref{eq:glo} can be further decomposed into the following problem:
\begin{align}
\label{eq:glo2}
\underset{\mathbf z_{\tilde j} \in \mathbf z_{\mathcal G_j} \setminus \mathbf z_{\mathbf s_j} }{\mymax}  I(\mathbf z_{\tilde j}; \theta)  &- \frac{1}{|\mathbf s|} \underset{\mathbf z_{\tilde t} \in \mathbf z_{\mathbf s_j}}{\sum} I (\mathbf z_{\tilde j}; \mathbf z_{\tilde t})  \\ \nonumber
&- \frac{1}{|\mathbf s|} \underset{\mathbf z_{t} \in \mathbf z_{\mathbf s} \setminus \mathbf z_{\mathbf s_j} }{\sum} I (\mathbf z_{\tilde j}; \mathbf z_{t}). \nonumber
\end{align}

For the second case, the problem in Equation \eqref{eq:glo} becomes
\begin{align}
\label{eq:glo3}
\underset{\mathbf z_{\tilde j} \in \cup_{\tilde i = i + 1}^{K} \mathbf z_{\mathcal G_{\tilde i}} }{\mymax}  I(\mathbf z_{\tilde j}; \theta)  -  \frac{1}{|\mathbf s|} \underset{\mathbf z_{t} \in \mathbf z_{\mathbf s}}{\sum} I (\mathbf z_{\tilde j}; \mathbf z_{t})  
\end{align}

Note that for the problems in Equation \eqref{eq:glo2} and Equation \eqref{eq:glo3}, we have $\mathbf z_t$ and $\mathbf z_{\tilde j}$ that are generated from different clusters. Using Assumption $A_2$, we have $P\left (d (\mathbf z_{\tilde j}, \mathbf z_t) > d_{th} \right) \geq 1 - \epsilon$, where $d_{th}, d_L < d_{th} < d_H$ is the threshold we used to cluster sensors. If the dissimilarity of two data sequences is greater than $d_{th}$, we put these sequences into two into different clusters; Otherwise, we put them into the same cluster. $\epsilon > 0$ is a small allowed tolerance. 
Furthermore, we assume that $I(\mathbf z_{\tilde j}; \mathbf z_t)$ 
is a non-increasing function of the dissimilarity $d (\mathbf z_{\tilde j}, \mathbf z_t)$. Based on  Assumption $A_2$, we have
\begin{equation}
P(I(\mathbf z_{\tilde j}; \mathbf z_t) < \zeta ) \geq 1 - \epsilon,
\end{equation}
where $\zeta$ is the obtained mutual information with dissimilarity $d_{th}$. Note that the empirical mutual information $I(\mathbf z_{\tilde j}; \mathbf z_t)$ also depends on the number of data samples that are available. In this proof, we assume that we have enough data samples to estimate the empirical mutual information accurately.

Note that the closed form expression for $\zeta$ is difficult to obtain due to the complicated relationship between the mutual information and rank-based dependence measure. The mutual information and the rank-based dependence measure (Spearman's $\rho$ or Kendall's $\tau$) can be connected using the copula based dependence measure. For random variables $x$ and $y$, the connection between mutual information and copula-based dependence measure is given as 
\begin{equation}
I(x; y) = \int_{[0, 1]^2} c(u, v) \log c(u, v) d_ud_v,
\end{equation}
where $c$ is the copula density function between variables $x$ and $y$. Also, $u = F(x)$ and $v = F(y)$, where $F(\cdot)$ is the CDF. 

The connections between the rank-based dependence measures (Kendall's $\tau$ and Spearman's $\rho$) and the copula-based dependence measure are given as
\begin{align*}
\tau(x, y) = 4 \int_{[0,1]^2} C(u, v) dC(u, v) - 1, \\
\rho(x, y) = 12 \int_{u} \int_v C(u, v) dudv - 3.
\end{align*}

The computation of $\zeta$ can be carried our using numerical differentiation and integration. However, if $x$ and $y$ follow Gaussian distributions and are linearly dependent, we have
\begin{equation}
I(x; y) = - \frac{1}{2} \log (1 - r^2),
\end{equation}
where $r = \text{corr}(x, y)$ is the Pearson correlation coefficient. If we define our dissimilarity as $d(\cdot, \cdot) =  \sqrt{1 - r^2}$, we have $r_{th}^2 = 1 - d_{th}^2$ given $d_{th}$. Therefore, $\zeta = - \log d_{th}$. As we can see that, $\zeta$ is a decreasing function of $d_{th}$.



In the following, our goal is to show that for the first case $P \left (\frac{1}{|\mathbf s|} \underset{\mathbf z_{t} \in \mathbf z_{\mathbf s} \setminus \mathbf z_{\mathbf s_j} }{\sum} I (\mathbf z_{\tilde j}; \mathbf z_{t})  > \zeta \right) < n_1 \epsilon$, where $1 \leq n_1 < L$, and for the second case, $P \left ( \frac{1}{|\mathbf s|} \underset{\mathbf z_{t} \in \mathbf z_{\mathbf s}}{\sum} I (\mathbf z_{\tilde j}; \mathbf z_{t}) > \zeta \right)  < n_2 \epsilon$, where $1 \leq n_2 < L$. 

We first prove for the first case.
\begin{align*}
&P \left (\frac{1}{|\mathbf s|} \underset{\mathbf z_{t} \in \mathbf z_{\mathbf s} \setminus \mathbf z_{\mathbf s_j} }{\sum} I (\mathbf z_{\tilde j}; \mathbf z_{t})  > \zeta \right) \\ \nonumber
&< P \left (\frac{1}{|\mathbf s \setminus \mathbf s_j |} \underset{\mathbf z_{t} \in \mathbf z_{\mathbf s} \setminus \mathbf z_{\mathbf s_j} }{\sum} I (\mathbf z_{\tilde j}; \mathbf z_{t})  > \zeta \right), \\ \nonumber
&< P \left (\underset{\mathbf z_{t} \in \mathbf z_{\mathbf s} \setminus \mathbf z_{\mathbf s_j} }{\max} I (\mathbf z_{\tilde j}; \mathbf z_{t}) \geq \zeta \right), \\ \nonumber
&= 1-\underset{\mathbf z_{t} \in \mathbf z_{\mathbf s} \setminus \mathbf z_{\mathbf s_j}}{\prod} P \left (I (\mathbf z_{\tilde j}; \mathbf z_{t}) < \zeta \right), \\ \nonumber
&= 1-(1-\epsilon)^{n_1} \\ \nonumber
&< n_1\epsilon, \nonumber
\end{align*}
where $n_1 =  |\mathbf s \setminus \mathbf s_j|$, and $1\leq n_1 \leq L$.

Therefore, we have $P \left (\frac{1}{|\mathbf s|} \underset{\mathbf z_{t} \in \mathbf z_{\mathbf s} \setminus \mathbf z_{\mathbf s_j} }{\sum} I (\mathbf z_{\tilde j}; \mathbf z_{t})  > \zeta \right) < n_1 \epsilon$, where $1 \leq n_1 < L$. Similarly, we can show that $P \left ( \frac{1}{|\mathbf s|} \underset{\mathbf z_{t} \in \mathbf z_{\mathbf s}}{\sum} I (\mathbf z_{\tilde j}; \mathbf z_{t}) > \zeta \right)  < n_2 \epsilon$, where $1 \leq n_2 < L$.

By suitably designing $d_{th}$, we can make $\zeta $ sufficiently small. Therefore, with probability at least $1 - \epsilon$, the term $\frac{1}{|\mathbf s|} \underset{\mathbf z_{t} \in \mathbf z_{\mathbf s} \setminus \mathbf z_{\mathbf s_j} }{\sum} I (\mathbf z_{\tilde j}; \mathbf z_{t})$ and the term $\frac{1}{|\mathbf s|} \underset{\mathbf z_{t} \in \mathbf z_{\mathbf s}}{\sum} I (\mathbf z_{\tilde j}; \mathbf z_{t})$ are upper bounded by  $\zeta$. 

For the first case, by ignoring the term $\frac{1}{|\mathbf s|} \underset{\mathbf z_{t} \in \mathbf z_{\mathbf s} \setminus \mathbf z_{\mathbf s_j} }{\sum} I (\mathbf z_{\tilde j}; \mathbf z_{t})$, the problem in Equation \eqref{eq:glo2} reduces to 
\begin{equation}
\label{eq:loc}
\underset{\mathbf z_{\tilde j} \in \mathbf z_{\mathcal G_j} \setminus \mathbf z_{\mathbf s_j} }{\mymax}  I(\mathbf z_{\tilde j}; \theta)  - \frac{1}{|\mathbf s|} \underset{\mathbf z_{\tilde t} \in \mathbf z_{\mathbf s_{j}} }{\sum} I (\mathbf z_{\tilde j} ; \mathbf z_{\tilde t}).
\end{equation}

Since $\frac{1}{|\mathbf s|}$ is a scale parameter, which will not affect the solution of the problem in Equation \eqref{eq:loc}, 
the above optimization problem can be further written as
\begin{equation}
\underset{\mathbf z_{\tilde j} \in \mathbf z_{\mathcal G_j} \setminus \mathbf z_{\mathbf s_{j}} }{\mymax}  I(\mathbf z_{\tilde j}; \theta)  - \frac{1}{|\mathbf s_{j}|} \underset{\mathbf z_{\tilde t} \in \mathbf z_{\mathbf s_{j}} }{\sum} I (\mathbf z_{\tilde j} ; \mathbf z_{\tilde t}),
\end{equation}
which is equivalent to the cluster-based incremental search problem in Equation \eqref{eq:incre}. 

For the second case, by ignoring the term $\frac{1}{|\mathbf s|} \underset{\mathbf z_{t} \in \mathbf z_{\mathbf s}}{\sum} I (\mathbf z_{\tilde j}; \mathbf z_{t})$, the problem in Equation \eqref{eq:glo3} reduces to cluster-based incremental search problem in Equation \eqref{eq:incre} while selecting the first sensor in the cluster.

\begin{remark}
Since we don't consider weakly dependent sensors (nearly independent sensors) within a cluster in this work, for the first case in Equation \eqref{eq:glo2}, the term $\frac{1}{|\mathbf s|} \underset{\mathbf z_{\tilde t} \in \mathbf z_{\mathbf s}}{\sum} I (\mathbf z_{\tilde j}; \mathbf z_{\tilde t})$ is significantly larger than the term $\frac{1}{|\mathbf s|} \underset{\mathbf z_{t} \in \mathbf z_{\mathbf s} \setminus \mathbf z_{\mathbf s_j} }{\sum} I (\mathbf z_{\tilde j}; \mathbf z_{t})$. The extreme scenario is that the term $\frac{1}{|\mathbf s|} \underset{\mathbf z_{\tilde t} \in \mathbf z_{\mathbf s}}{\sum} I (\mathbf z_{\tilde j}; \mathbf z_{\tilde t})$ is a small number due to a large scale parameter $|\mathbf s|$.  For this scenario, the dominant term would be $I(\mathbf z_{\tilde j}; \theta)$ which can be covered by the second case in Equation \eqref{eq:glo3}.
\end{remark}

Therefore, by designing $d_{th}$, with at least probability $1 - \epsilon$, the global incremental search method in Equation \eqref{eq:increO} reduces to cluster-based incremental search. \hfill $\blacksquare$

\bibliographystyle{IEEE}
\bibliography{consensus_estimation,refcopulaJ,reff_shan}

\end{document}